\documentclass{emulateapj}

\usepackage{graphicx, subfigure}
\usepackage{amssymb}
\usepackage{amsmath}
\usepackage{epstopdf}
 \usepackage{hyperref}
\usepackage{color}
\usepackage{placeins}
\usepackage{natbib}
\usepackage{enumerate}
\usepackage{textcomp}
\usepackage{footmisc}
\usepackage{xspace}

\newcommand{\Fermi}{\emph{Fermi}\xspace}
\newcommand{\fermi}{\emph{Fermi}\xspace}
\newcommand{\fraz}[2]{\frac{\displaystyle #1}{\displaystyle #2}}

\def\Rs{$R_\odot$}
\def\p0{$\pi^{\rm 0}$}

\def\de{$^{\circ}$\xspace}
\def\e{\epsilon}

\def\like{\mathcal{L}}
\def\E{\mathcal{E}}

\newcommand{\nob}[1]{{ #1}}
\shorttitle{The bright solar flare of March 2012}
\shortauthors{\Fermi-LAT Collaboration}
\begin{document}

\title{Impulsive and Long Duration High-Energy Gamma-ray Emission From the Very Bright 2012 March 7 Solar Flares}
\author{
M.~Ajello\altaffilmark{2}, 
A.~Albert\altaffilmark{3}, 
A.~Allafort\altaffilmark{4}, 
L.~Baldini\altaffilmark{5}, 
G.~Barbiellini\altaffilmark{6,7}, 
D.~Bastieri\altaffilmark{8,9}, 
R.~Bellazzini\altaffilmark{10}, 
E.~Bissaldi\altaffilmark{11}, 
E.~Bonamente\altaffilmark{12,13}, 
T.~J.~Brandt\altaffilmark{14}
% ...
}
% ...
\altaffiltext{2}{Space Sciences Laboratory, 7 Gauss Way, University of California, Berkeley, CA 94720-7450, USA}
\altaffiltext{3}{Department of Physics, Center for Cosmology and Astro-Particle Physics, The Ohio State University, Columbus, OH 43210, USA}
\altaffiltext{4}{W. W. Hansen Experimental Physics Laboratory, Kavli Institute for Particle Astrophysics and Cosmology, Department of Physics and SLAC National Accelerator Laboratory, Stanford University, Stanford, CA 94305, USA}
\altaffiltext{5}{Universit\`a  di Pisa and Istituto Nazionale di Fisica Nucleare, Sezione di Pisa I-56127 Pisa, Italy}
\altaffiltext{6}{Istituto Nazionale di Fisica Nucleare, Sezione di Trieste, I-34127 Trieste, Italy}
\altaffiltext{7}{Dipartimento di Fisica, Universit\`a di Trieste, I-34127 Trieste, Italy}
\altaffiltext{8}{Istituto Nazionale di Fisica Nucleare, Sezione di Padova, I-35131 Padova, Italy}
\altaffiltext{9}{Dipartimento di Fisica e Astronomia ``G. Galilei'', Universit\`a di Padova, I-35131 Padova, Italy}
\altaffiltext{10}{Istituto Nazionale di Fisica Nucleare, Sezione di Pisa, I-56127 Pisa, Italy}
\altaffiltext{11}{Institut f\"ur Astro- und Teilchenphysik and Institut f\"ur Theoretische Physik, Leopold-Franzens-Universit\"at Innsbruck, A-6020 Innsbruck, Austria}
\altaffiltext{12}{Istituto Nazionale di Fisica Nucleare, Sezione di Perugia, I-06123 Perugia, Italy}
\altaffiltext{13}{Dipartimento di Fisica, Universit\`a degli Studi di Perugia, I-06123 Perugia, Italy}
\altaffiltext{14}{NASA Goddard Space Flight Center, Greenbelt, MD 20771, USA}
% ...
\begin{abstract}
%%\date{\today}
The \Fermi Large Area Telescope (LAT) observed two bright X-class solar flares on 2012 March 7, and detected gamma-rays up to 4 GeV. We detected gamma-rays both during the impulsive and temporally-extended emission phases, with emission above 100 MeV lasting for approximately 20 hours. Accurate localization of the gamma-ray production site(s) coincide with the solar active region from which X-ray emissions associated with these flares originated. Our analysis of the $>100$ MeV gamma-ray emission shows a relatively rapid monotonic decrease in flux during the first hour of the impulsive phase, and a much slower, almost monotonic decrease in flux for the next 20 hours. The spectra can be adequately described by a power law with a high energy exponential cutoff, or as resulting from the decay of neutral pions produced by accelerated protons and ions with an isotropic power-law energy distribution. 
The required proton spectrum has a number index $\sim 3$, with minor variations during the impulsive phase, while during the temporally extended phase the spectrum softens monotonically, starting with index $\sim 4$. 
The $>30$ MeV proton  flux and spectra observed near the Earth by the GOES satellites also show a monotonic flux decrease and spectral softening during the extended phase, but with a harder spectrum, with index $\sim 3$. Based on the \Fermi-LAT and GOES observations of the flux and spectral evolution of these bright flares, we explore the relative merits of prompt and continuous acceleration scenarios, hadronic and leptonic emission processes, and acceleration at the solar corona by the fast Coronal Mass Ejections (CME) as explanations for the observations. We conclude that the most likely scenario is continuous acceleration of protons in the solar corona which penetrate the lower solar atmosphere and produce pions that decay into gamma-rays.  
\end{abstract}


\section{Introduction}

Measurements of the long lasting gamma-ray emission from bright solar flares 
provide the opportunity to investigate the impulsive energy release and 
acceleration mechanisms responsible for these explosive phenomena. The June 
11 1991 Geostationary Operational Satellite Server (GOES) X12.0 class 
flare observed by the EGRET instrument on-board the Compton Gamma-ray 
Observatory \nob{\citep{Hughes:80,Kanbach:88,Thompson:93,Esposito:99}} produced
gamma-rays with energies greater than 100 MeV up to 8 hours after the 
impulsive phase, setting a 
record for the detection of long lasting emission of high-energy photons 
\citep{kanb93}. The origin of this temporally extended emission is not well 
understood. Important questions such as whether (i) the radiative process is 
hadronic or leptonic, (ii) the acceleration happens at the flare site or at 
the Coronal Mass Ejection (CME), (iii) continuous acceleration or trapping and 
precipitation are required, are still debated. Additional and more detailed 
flare observations are clearly necessary to fully understand 
the mechanisms at work to produce the high-energy gamma-rays.

The \Fermi observatory is comprised of two instruments: the Large Area 
Telescope (LAT) designed to detect gamma rays from 20~MeV up to more than 
300 GeV~\citep{LATPaper} and the Gamma-ray Burst Monitor (GBM) which is 
sensitive from $\sim 8$ keV up to 40~MeV \citep{GBMinstrument}. The orbital
inclination of the \Fermi satellite is 25.6$^{\circ}$ with an altitude of 565 km
and completes one full orbit every 90 minutes.

The LAT has already detected several flares above 100 MeV, during both the 
impulsive and the temporally extended phases \nob{\citep{2011ATel.3635....1O,2011ATel.3552....1O,2012ATel.3886....1T,2012AAS...22042404P,2012AAS...22042403O}}. 
The first \Fermi GBM and LAT detection from the impulsive GOES M2.0 
flare of June 12 2010 is presented in~\citet{2012ApJ...745..144A}.
The analysis of this flare was performed using the LAT Low-Energy (LLE) 
technique \nob{(see Appendix \ref{sec:LLE})} because the soft X-rays emitted during the prompt 
emission of a flare penetrate the anti-coincidence detector (ACD) of the LAT 
causing a pile-up effect which can result in a significant decrease in 
gamma-ray detection efficiency in the standard on-ground photon analysis. The pile-up 
effect has been addressed in detail and we refer the reader 
to \citet{2012ApJ...745..144A} and \citet{LATperform} for a full description.
The list of other flares, and the analysis of the first two flares with long lasting high-energy emission (March 7--8 2011 and June 7 2011) is presented in (\Fermi-LAT collaboration, in preparation).

Here we report on impulsive and long-duration high-energy gamma-ray emission observed by \Fermi LAT and associated with the intense X-ray solar flares of 2012 March 7. In the next section (\S\,2) we present the temporal evolution of soft X-ray and Solar Energetic Particles (SEP) fluxes as measured by GOES; in \S\,3
we describe the details of  the gamma-ray analysis, and finally, in \S\,4 we discuss and interpret our results.
%In the next section (\S\,2) we present the temporal evolution of soft X-ray and Solar Energetic Particles (SEP) fluxes as measured by GOES; in \S\,3
%we describe the localization of the high-energy gamma-ray emission on the solar disk, in \S\,4 we describe the details of 
%the time-resolved analysis of LAT data for the whole period of the observation, and finally, in \S\,5 we discuss and interpret our results.

\section{GOES X-ray and Solar Energetic Particles}
On 2012 March 7 two bright X-class flares originating from the active region NOAA AR\#:11429 (located at N16E30) erupted within an hour of each other, marking one of the most active days of Solar Cycle 24. 
The first flare started at 00:02:00 UT and reached its maximum intensity (X5.4) at 00:24:00 UT while the second X1.3 class flare occurred at 01:05:00 UT, reaching its maximum 9 minutes later. 

The GOES satellite observed intense X-ray emission beginning at about 00:05:00 UT and lasting for several hours. Moreover, it detected in three energy bands Solar Energetic Particle (SEP) protons originating from these flares. 
In the top panels of Figures~\ref{lightcurve1} and \ref{lightcurve2} the GOES X-ray data measured in both 3--25 keV and 1.5--12 keV channels are shown for two time intervals during the flaring episode. GOES soft X-ray light curves usually do not follow the impulsive nature of the activity because they trace the accumulated energy input by the accelerated particles. In general, based on the so-called \emph{Neupert effect} \citep{1968ApJ...153L..59N} the derivative of these light curves is considered to be a good proxy for the temporal evolution of the accelerated-particle interactions. Figure~\ref{lightcurve1} shows the light curves for the 1.5--12 and 3--25 keV GOES bands, together with their corresponding derivatives. Such derivatives make it clear that the first flare consisted of two impulsive bursts with a duration of a few minutes each while the second flare was composed of only one such pulse.
In the top panel of Figure~\ref{lightcurve2}, we display the 5-minutes average rate of protons detected by the GOES satellite in three energy bands (30--50 MeV, 50--100 MeV and $>$100 MeV).
Unfortunately the Reuven Ramaty High-Energy Solar Spectroscopic Imager \citep[RHESSI,][]{2002SoPh..210....3L} was not observing the Sun during this period.

%\nob{X-ray data from GOES satellite 15 measured in in the short (0.05--0.4\,nm) and the long (0.1--0.8\,nm) wavelength channel
% are shown in the top panels of Figure~\ref{lightcurve1} and \ref{lightcurve2}, for two different time window.}

\section{\Fermi gamma-ray data}
Orbital sunrise for \Fermi occurred less than six minutes after the peak of the first flare, triggering the GBM at 00:30:32.129 UT 
(causing the abrupt rate increase visible in Figure\ref{lightcurve1}, middle panel).
The second flare is also clearly visible in the BGO$_{0}$ detector of the GBM\footnote{We use dead-time corrected count rates from the NaI$_{2}$ in the 10--25\,keV and 100--300\,keV energy bands and from the BGO$_{0}$ in the 1--10\,MeV energy band.}.
The \Fermi LAT $>$100 MeV count rate\footnote{for \texttt{P7SOURCE\_V6} class with a cut on the zenith angle, z$_{max}$=100$^{\circ}$} was dominated by the gamma-ray emission from the Sun\footnote{http://apod.nasa.gov/apod/ap120315.html}, which was nearly 100 times brighter than the Vela Pulsar in the same energy range.
During the impulsive phase (the first eighty minutes) the X5.4 flare was so intense that the LAT anti-coincidence detector suffered from pulse pileup(see Appendix \ref{appendix::LLE}), so the standard instrument response functions (IRFs) could not be used. Instead, the spectral analysis performed for the first 80 minutes differs from that used at later times; moreover, we exclude the impulsive phase from the localization analysis.

To limit the possible bias due to the so-called ``fisheye'' effect \citep[][\S 6.4]{2012ApJS..203....4A} we used the true 
position of the Sun when it was in the field of view of the LAT to calculate an energy and angle-dependent correction that was applied to the reconstructed photon direction on an event-by-event basis.

\subsection{The impulsive phase}
\label{sec:LLE}
\begin{figure*}[ht!]
\begin{center}
\includegraphics[width=\linewidth]{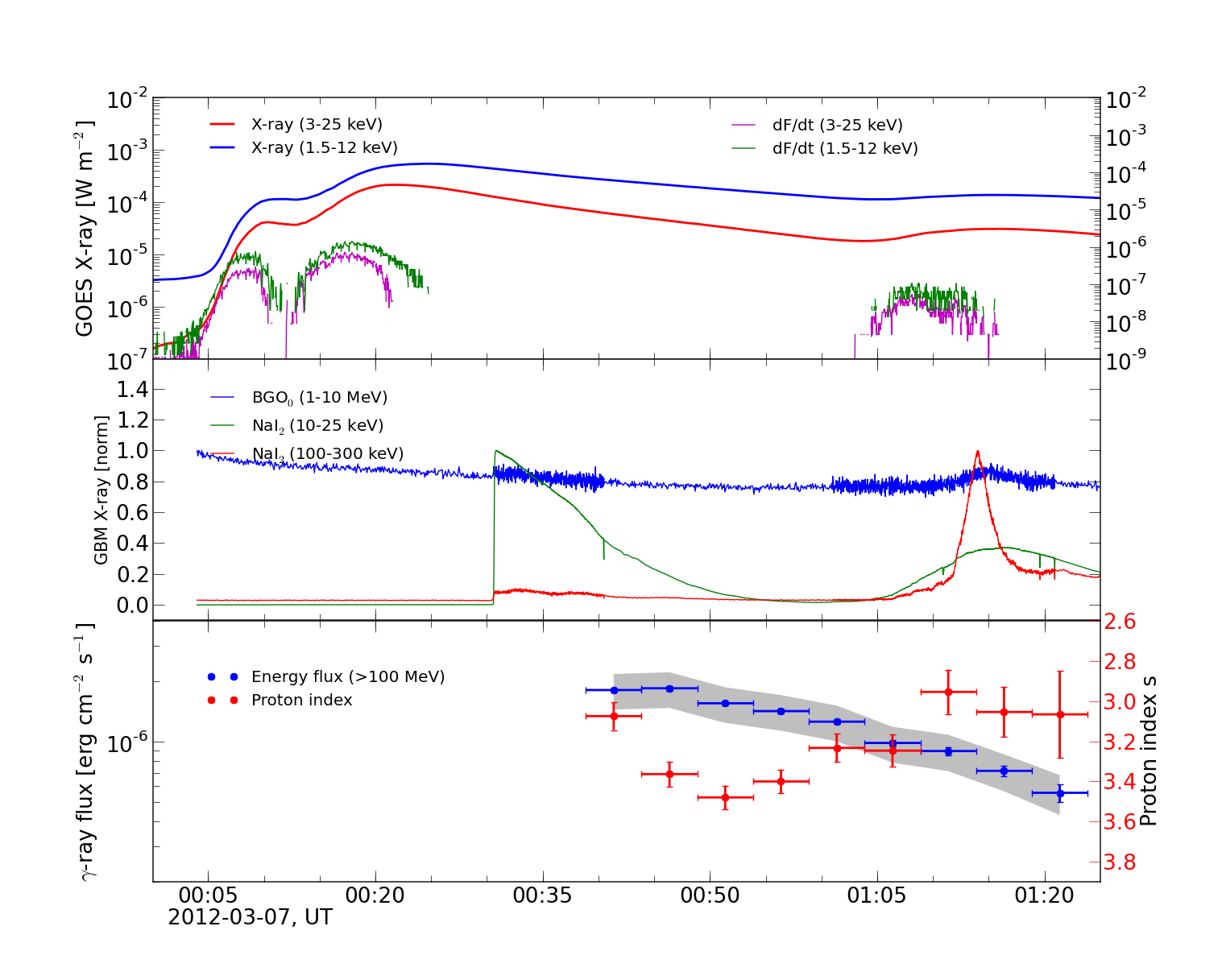}
\caption{Composite light curves for 2012 March 7 flare, covering the first $\sim$80 minutes. 
{\bf Top panel}: Soft X-rays (red:\,1.5--12 keV, blue:\,3--25 keV) from the GOES 15 satellite. On the right axis are the first derivatives of the soft X-rays fluxes (magenta:\,1.5--12 keV, green:\,3--25 keV).
These curves approximate accelerated electron impulsive lightcurves \citep{1968ApJ...153L..59N}.
{\bf Middle panel}: Hard X-rays count rates from the GBM; green and red for NaI$_{2}$ 10--25\,keV and 100--300\,keV energy channels, and blue for the BGO$_{0}$ detector. 
{\bf Bottom panel}: LAT ($>$100 MeV) gamma-ray flux (blue) and derived proton spectral index (red). 
The gray band represents the systematic uncertainties associated to the flux measurement, and it is obtained by adding 20\% systematic error in quadrature.}
\label{lightcurve1}
\end{center}
\end{figure*}

\begin{figure*}[ht!]
\begin{center}
\includegraphics[width=\linewidth]{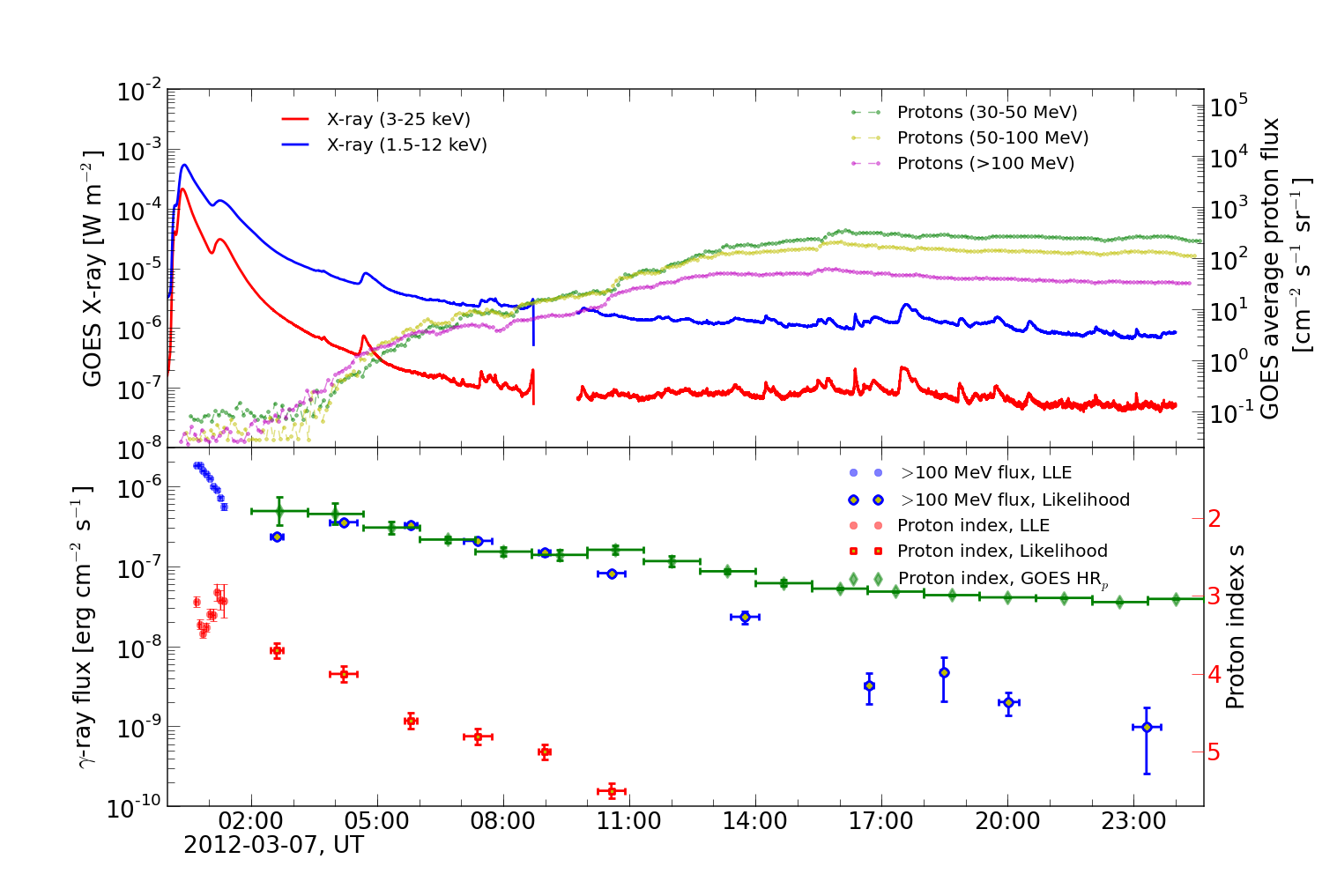}
\caption{Long lasting emission. {\bf Top panel}: soft X-rays (red: 1.5--12 keV, blue: 3--25 keV) from the GOES 15 satellite. On the right axis, 5-minute averaged proton flux (green: 30--50 MeV, yellow: 50--100 MeV, magenta: $>$100 MeV). We display the average of detectors A and B.
{\bf Bottom panel}: high energy gamma ray flux above 100 MeV measured by the \Fermi LAT. The Blue/red circles represent the flux and the derived proton spectral index obatined with the LLE analysis (covering the initial period, when the instrumental performance was affected by pileup of hard-X-rays in the ACD tiles). Blue/red empty circles/squares represent the flux and the derived proton spectral index obtained by standard likelihood analysis. Green diamonds are the GOES proton spectral indexes derived from the hardness ratio, as described in the text.}
\label{lightcurve2}
\end{center}
\end{figure*}

The first step in our analysis of the impulsive phase is to consider 9 adjacent time intervals of LLE data (see Appendix \ref{appendix::LLE} for a detailed description of the LLE technique). In particular, we included data for the region centered on the Sun at the time corresponding to the middle of each time interval and selected the intervals when the Sun was within 70$^{\circ}$ from the LAT boresight. For each time interval, we extract two sets of background LLE data at 30 orbits before and after the flare, when \Fermi was at a similar geomagnetic location. At $\pm$30 orbits ($\sim$2 days) the location and attitude of the spacecraft are approximately the same as during the impulsive phase of the flare. To compensate for the rotation of the Earth, 
the center of the region of interest (ROI) is held fixed in instrument coordinates \nob{at the center of each time bin}. This last step is needed to average the two background data sets because the local cosmic-ray-induced background dominates the in-aperture celestial background. In this way we obtain 9 source spectra and 9 background spectra (one for each time interval \nob{during the impulsive phase analysis}). We compute the LLE energy redistribution matrices for each of the 9 intervals separately.

We fit the data between 100~MeV and 10~GeV using \texttt{XSPEC}\footnote{http://heasarc.gsfc.nasa.gov/docs/xanadu/xspec/index.html} to test three models. The first two are simple phenomenological functions, to describe bremsstrahlung emission from accelerated electrons, namely a pure power law (PL) and a power
law with an exponential cut-off (PLEXP):
\begin{equation}
 \frac{dN(E)}{dE} = N_{0}\,\e^{-\Gamma}\,\exp\left({-\frac{E}{E_{co}}}\right);
\label{eq1}
\end{equation} 
where $\Gamma$ is the photon index and $E_{co}$ is the cut-off energy.
We found that the data clearly diverge from a pure power law spectrum and that the PLEXP provides a better fit in all time intervals considered.
The third model uses templates based on a detailed study of the 
gamma rays produced from pion decay~\citep{murp87}. 
In this model accelerated high-energy protons (and other ions) with an assumed energy distribution collide with particles of the solar atmosphere, creating $\pi^0$ and $\pi^\pm$ mesons. A $\pi^0$ quickly decays into two gamma-rays, each having an energy of 67.5 MeV in the rest frame of the meson. The
decay products of charged $\pi^\pm$ mesons (secondary e$^\pm$) produce 
gamma-rays via Bremsstrahlung or by annihilation-in-flight of the positrons, and microwaves via synchrotron radiation\footnote{The interactions between the accelerated and background protons (and ions) also produce nuclear de-excitation lines in the 1 to 10 MeV range, observable by the GBM. The analysis of these flaring episodes by the GBM will be presented in a subsequent paper.}.
The ratio of the e$^{\pm}$ energy going to gamma-rays or to microwaves is $\sim (n/10^{14}
{\rm cm}^{-3})(1000{\rm G}{/B)}^2({\rm GeV}/E)$.  For the magnetic field strength in the solar atmosphere $B=300$ G used in our templates, synchrotron losses have only a small effect on the transport of the protons to column depths of $>10^{22}$ cm$^{-2}$ (needed to stop them) or densities of $n>10^{15}$\,cm$^{-3}$. However, the synchrotron emission may be detectable (see discussion below).

The pion-decay templates used in our fits depends on the ambient
density, composition and magnetic field, on the accelerated-particle composition, pitch angle distribution and energy spectrum. 
The templates represent a particle population with an isotropic pitch angle distribution and a power-law energy spectrum ($dN/dE \propto E^{-s}$, with $E$ the kinetic energy of the protons) 
interacting in a thick target with a coronal composition~\citep{ream95} taking 
$^4$He/H = 0.1. %\nob{The value of the magnetic field is 300~G.}
To obtain the gamma-ray flux value we fit the data varying the proton spectral index $s$ from 
2--6, in steps of 0.1. In this way, we fit the LAT data 
with a model with two free parameters, the normalization and the proton index, $s$.
The time dependence of the $>$100\,MeV gamma-ray flux and of the proton index, $s$, derived using gamma-ray LAT data, is displayed in the lower panel of Figure~\ref{lightcurve1} and the numerical values are reported in Table \ref{SpectralFit1}, as well as the best-fit parameters of the PLEXP model.

It appears that, after a short phase of spectral softening, the proton spectral index hardens before the start of the impulsive phase of the second flare as seen by the GBM detectors (middle panel of Figure~\ref{lightcurve1}). The spectral index $s$ correlates better with the GBM flux than with the high-energy flux measured by the LAT. For the interpretation of these results, see \S 4.

%To complete our time resolved spectral analysis we also study the behavior
%of the proton and X-ray flux in relation to the gamma-ray flux during the 
%flares. We used the proton flux from 
%the GOES satellite 13. In particular the 5-minutes 
%average flux with energies greater than 30, 50 and 100 MeV 
%from detectors A and B that faces either East or West depending on the yaw 
%flip of the satellite with a correction applied to remove contaminating 
%particles.
%We display the proton's fluxes obtained by differentiating the average fluxes in consecutive bands, 
%obtaining the fluxes between 30 and 50 MeV, 50 and 100 MeV and $>$ 100 MeV.

\begin{deluxetable*}{rcrrlr}
\tabletypesize{\scriptsize}
\tablecaption{Spectral analysis of the impulsive phase}
\tablewidth{0pt}
\tablehead{ 
\colhead{Time Interval} & \colhead{Proton index} & \colhead{Energy Flux$^{a}$} &\colhead{$\Gamma$} & \colhead{E$_{CO}$} & \colhead{Flux$^{a}$} \\
\colhead{2012/03/07 UT}  & \colhead{} & \colhead{} &\colhead{}    & \colhead{MeV}                    & \colhead{}}
\startdata
00:38:52--00:43:52  &  3.07$\pm$0.07 &    21$\pm$1    & $-$0.07$\pm$0.09 &  130$\pm$8 & 18.0$\pm$0.4 \\
00:43:52--00:48:52  &  3.36$\pm$0.07 &  18.7$\pm$0.6 & $-$0.26$\pm$0.07 & 107$\pm$4   & 16.3$\pm$0.3\\ 
00:48:52--00:53:52  &  3.48$\pm$0.07 & 15.5$\pm$0.3  & $-$0.23$\pm$0.06 & 106$\pm$4   & 14.1$\pm$0.2 \\ 
00:53:52--00:58:52  &  3.40$\pm$0.06 & 14.4$\pm$0.4 & $-$0.19$\pm$0.06 & 109$\pm$4   & 12.7$\pm$0.2 \\
00:58:52--01:03:52  &  3.23$\pm$0.07 & 12.7$\pm$0.4  & $-$0.18$\pm$0.07 & 114$\pm$5  & 10.9$\pm$0.2 \\
01:03:52--01:08:52  &  3.25$\pm$0.08 &   9.6$\pm$0.3 & $-$0.25$\pm$0.08 & 111$\pm$6   & 8.6$\pm$0.2 \\
01:08:52--01:13:52  &  2.95$\pm$0.08 &   9.0$\pm$0.3 &  0.00$\pm$0.05 & 136$\pm$7   & 7.2$\pm$0.2 \\
01:13:52--01:18:52  &  3.0$\pm$0.1   &   7.2$\pm$0.4 & 0.6$\pm$0.10      &  220$\pm$30 & 6.0$\pm$0.3 \\ 
01:18:52--01:23:52  &  3.1$\pm$0.2   &   5.7$\pm$0.6 & 0.8$\pm$0.20      &  270$\pm$70   & 5.0$\pm$0.5 \\ 
\enddata
\label{SpectralFit1}
\tablenotetext{a}{Integral energy flux between 100\,MeV and 10\,GeV, in units of $10^{-7}$ erg\,cm$^{-2}$\,s$^{-1}$.}
\end{deluxetable*}

%In Figure \ref{lightcurve1} we report the X-ray flux and the flux of 
%high-energy protons at Earth (top panel), the hard X-ray and soft 
%gamma-rays fluxes (middle panel) and the $>$100\,MeV gamma-ray flux 
%(bottom panel) for the first orbit. The red points in the bottom panel 
%represent the proton index $s$, derived using $\g$-ray LAT data.
%The soft X-ray light curve shows a first flare at approximately 
%00:10 UT followed by the intense X5.4 flare, at approximately 00:24 UT, while 
%no obvious feature can be seen in the proton spectrum for the entire time 
%covered by the x-axis (i.e., up to 1:25:00 UT). 

\subsection{Temporally extended emission}

Following the first 90 m of \Fermi-LAT observation up to the end of the flaring episode we perform our study using the standard likelihood analysis 
\nob{implemented in the \Fermi-LAT \texttt{ScienceTools}\footnote{We used \texttt{ScienceTools} version 09-28-00}} with \texttt{P7SOURCE\_V6} IRFs, selecting a 12\de radius ROI and selecting only photons that arrive at the LAT within 100 degrees of the zenith to reduce contamination from the Earth's limb.\nob{We include the azimuthal dependence of the effective area.}

To study the temporally extended emission, we perform time resolved spectral analysis in Sun-centered coordinates by transforming the reference system from celestial 
coordinates to ecliptic Sun-centered coordinates. This is necessary in order
to compensate for the effect of the apparent motion of the Sun during the long 
duration of the flare. We select intervals when the Sun was 
in the field of view (angular distance from the LAT boresight $<$ 70$^{\circ}$)
and use the unbinned maximum likelihood algorithm \texttt{gtlike}. 
We include the isotropic template model that is used to describe the extragalactic gamma-ray emission and the 
residual CR contamination (\texttt{iso\_p7v6source.txt}), leaving its normalization as the free parameter.
Over short time scales, the diffuse Galactic emissions produced by cosmic rays interacting with the interstellar medium are not spatially resolved and are hence included in the isotropic template.
We also add the gamma-ray emission from the quiescent Sun modeled as a point source located at the center of the disk, with a spectrum described with a simple power-law with a spectral index of 2.11 and an integrated flux ($>$ 100 MeV) of 4.6$\times$10$^{-7}$ ph\,cm$^{-2}$\,s$^{-1}$ \citep{2011ApJ...734..116A}.
\nob{We did not include the extended IC component described in \citet{2011ApJ...734..116A} because it is too faint to be detected during these time scales.}
We fit the data with the same two phenomenological functions used for the impulsive phase of the flare and use the Likelihood Ratio Test to estimate whether the addition of the exponential cut-off is statistically significant.
%We use the same three models to fit the data we used in the previous analysis.
%We use the Likelihood ratio test to estimate if the addition of the exponential cut-off is statistically significant.
The Test Statistic TS$=-2\,\Delta\log(\like)$ is twice the increment of the logarithm of the likelihood value $\like$ obtained by fitting the data adding the source to the background.
Because the null hypothesis is the same for the two cases, the increment of the Test Statistic ($\Delta$\,TS=TS$_{\rm PLEXP}$-TS$_{\rm PL}$) is equivalent to the corresponding difference of maximum likelihoods computed between the two models. 

\begin{figure*}[ht!]
\begin{center}
\includegraphics[trim=0cm 0cm 2cm 1cm, clip=true, width=3.5cm]{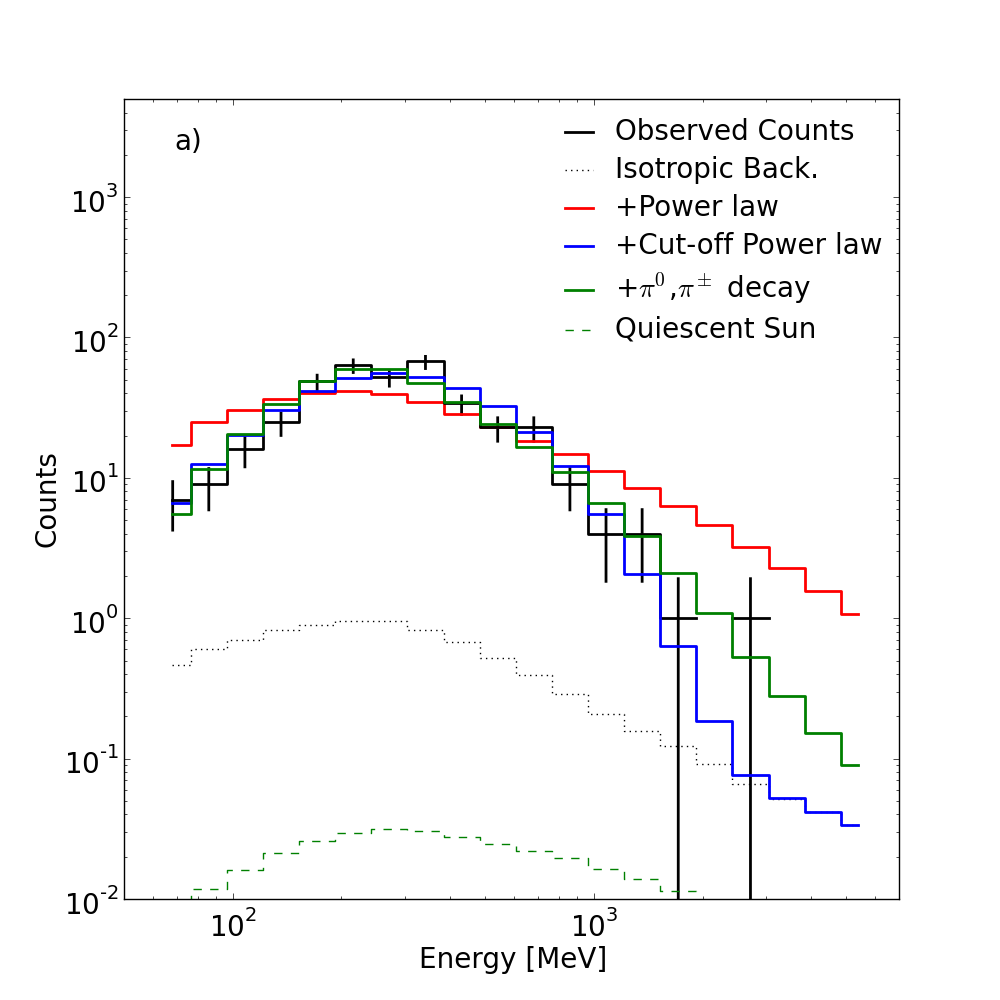}
\includegraphics[trim=0cm 0cm 2cm 1cm, clip=true, width=3.5cm]{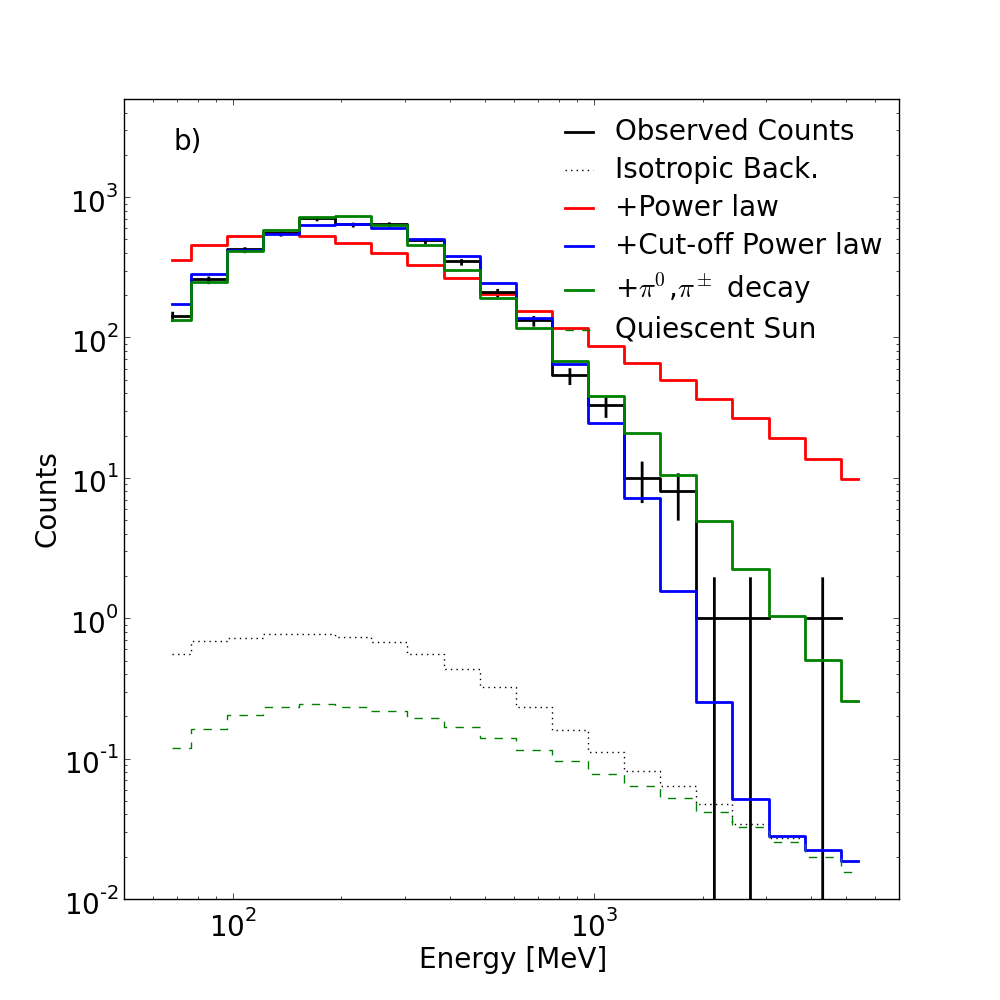}
\includegraphics[trim=0cm 0cm 2cm 1cm, clip=true, width=3.5cm]{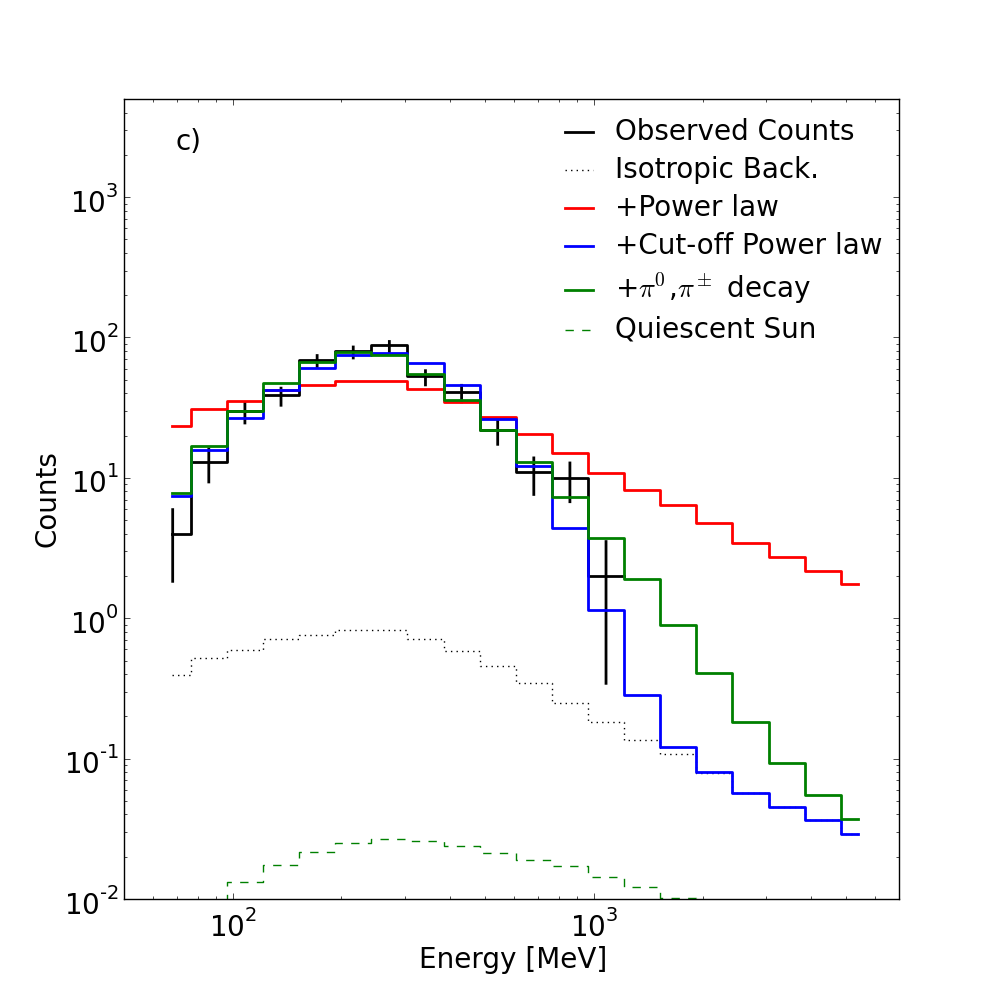}
\includegraphics[trim=0cm 0cm 2cm 1cm, clip=true, width=3.5cm]{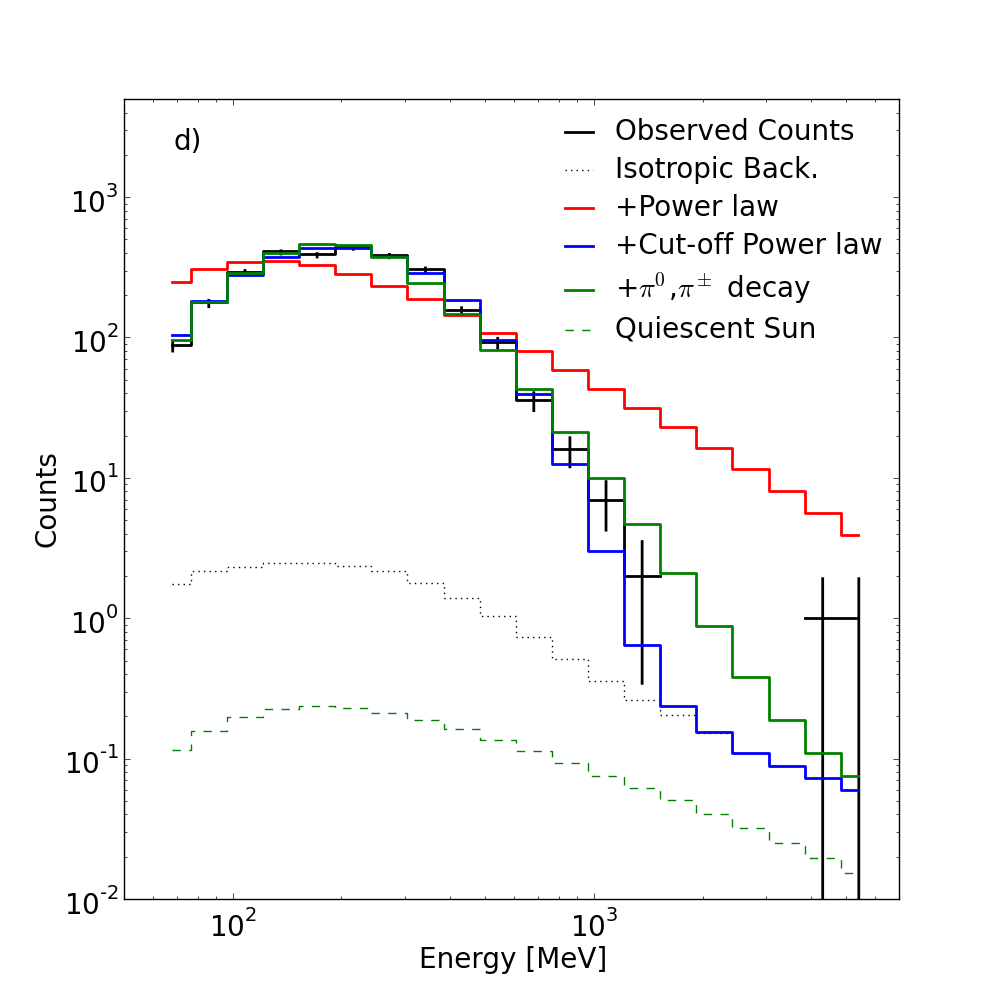}
\includegraphics[trim=0cm 0cm 2cm 1cm, clip=true, width=3.5cm]{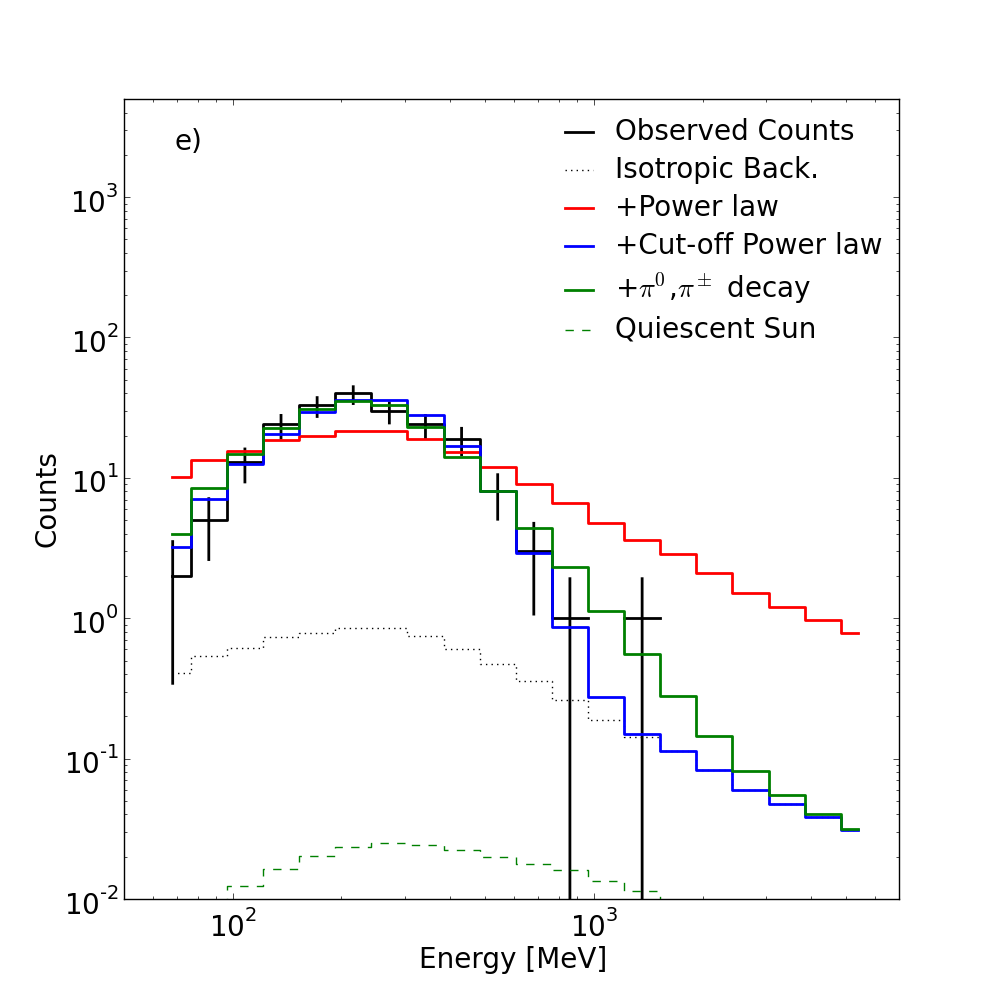}
\includegraphics[trim=0cm 0cm 2cm 1cm, clip=true, width=3.5cm]{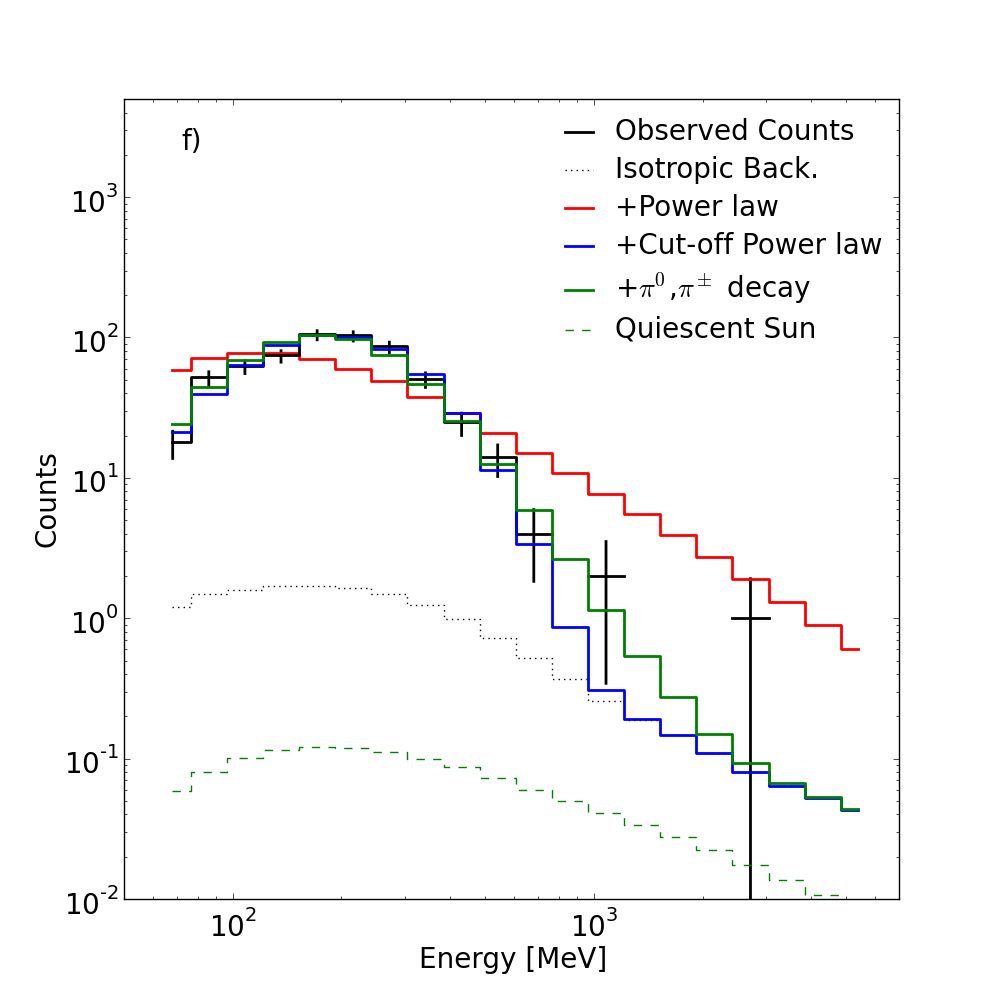}
\includegraphics[trim=0cm 0cm 2cm 1cm, clip=true, width=3.5cm]{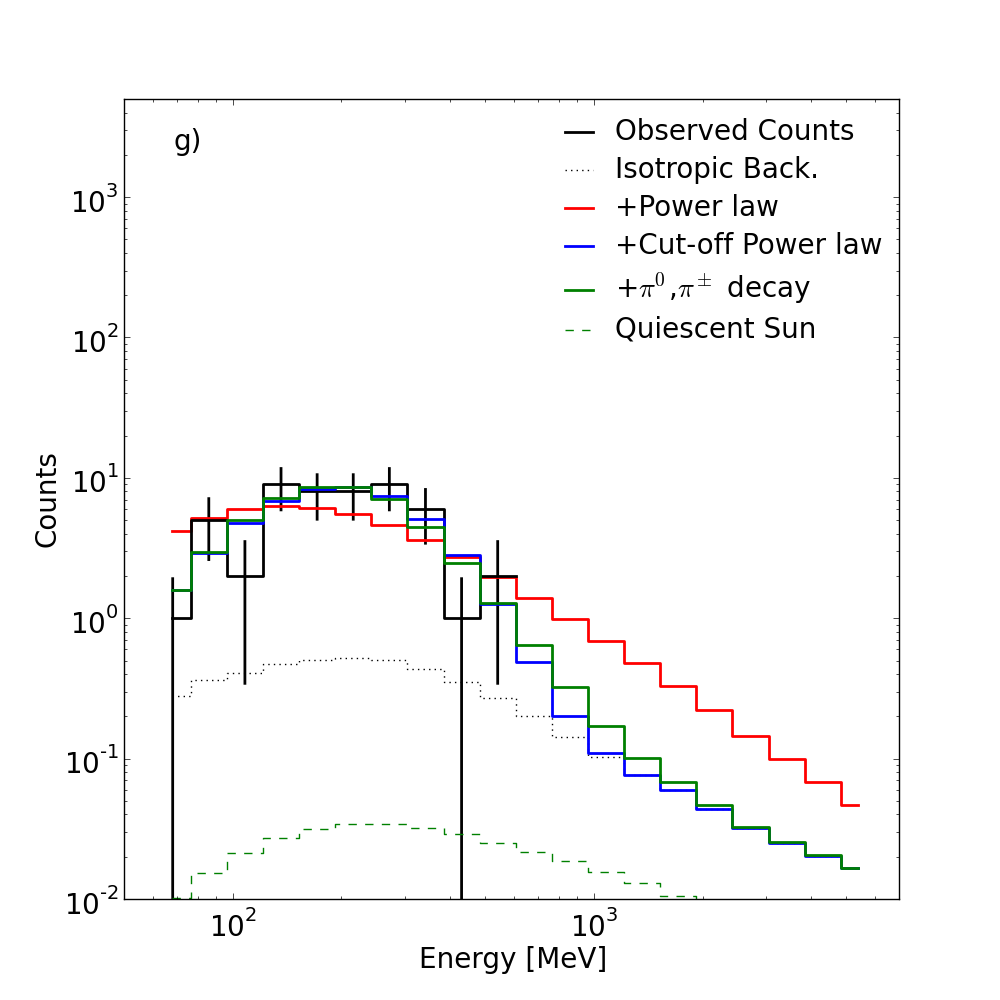}
\includegraphics[trim=0cm 0cm 2cm 1cm, clip=true, width=3.5cm]{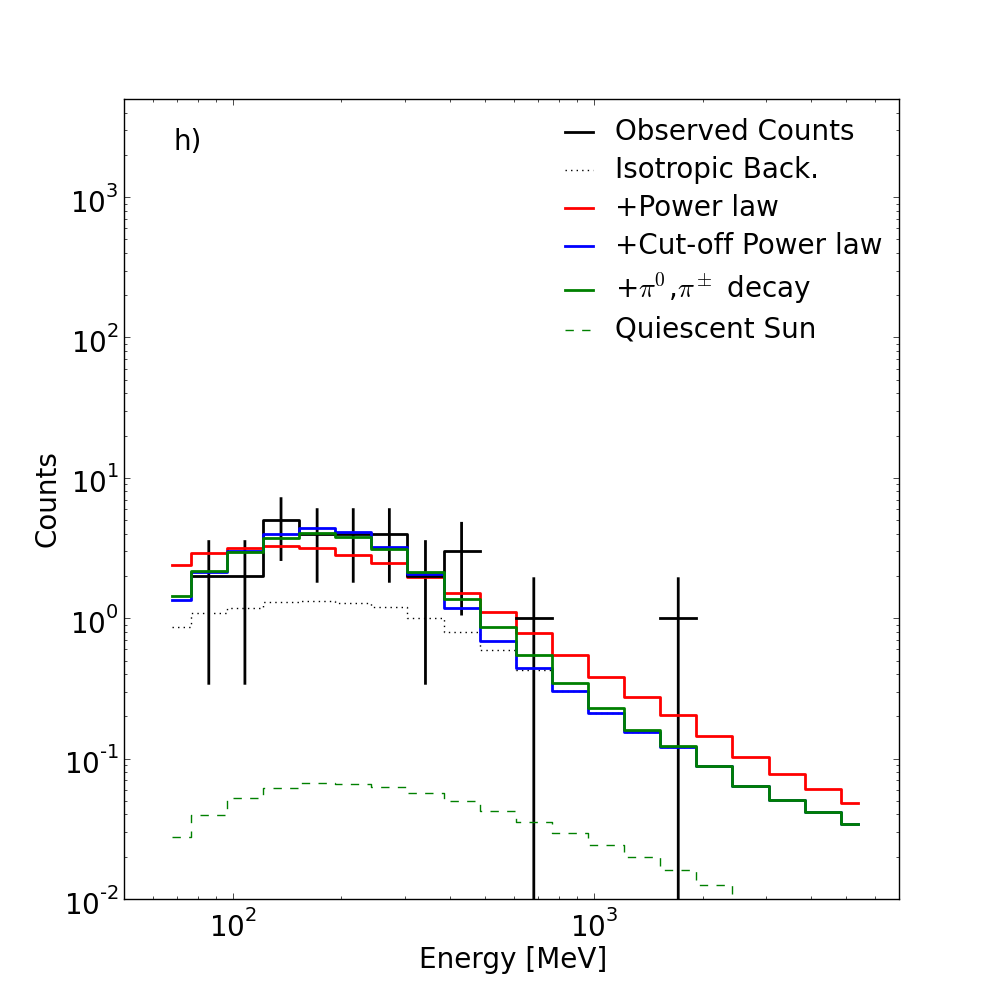}
\includegraphics[trim=0cm 0cm 2cm 1cm, clip=true, width=3.5cm]{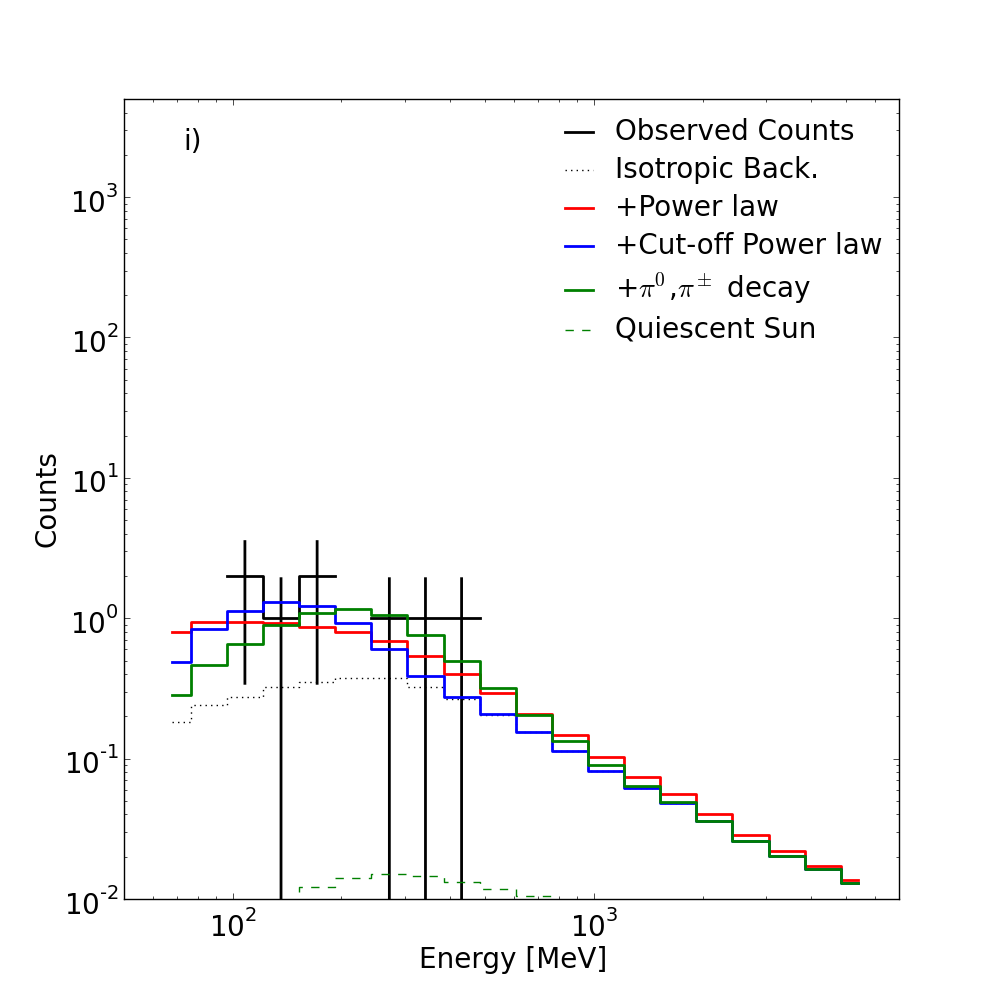}
\includegraphics[trim=0cm 0cm 2cm 1cm, clip=true, width=3.5cm]{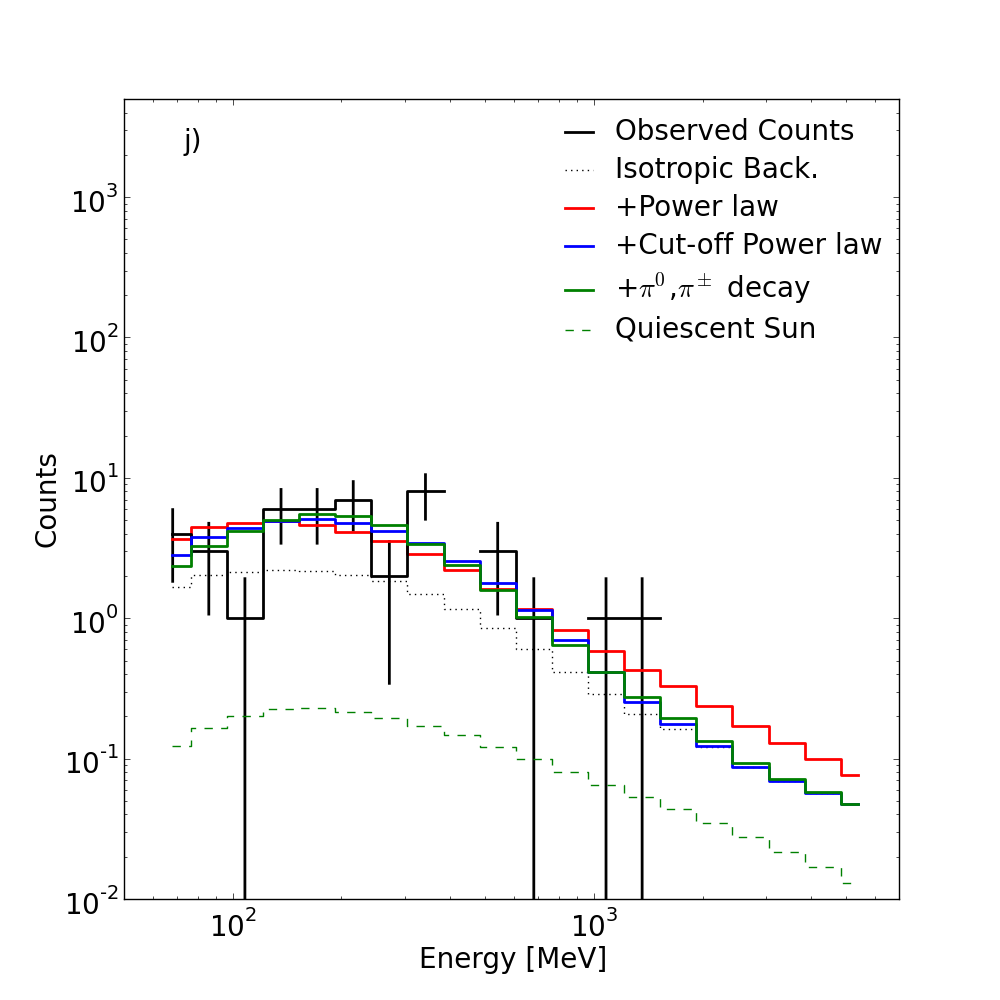}
\includegraphics[trim=0cm 0cm 2cm 1cm, clip=true, width=3.5cm]{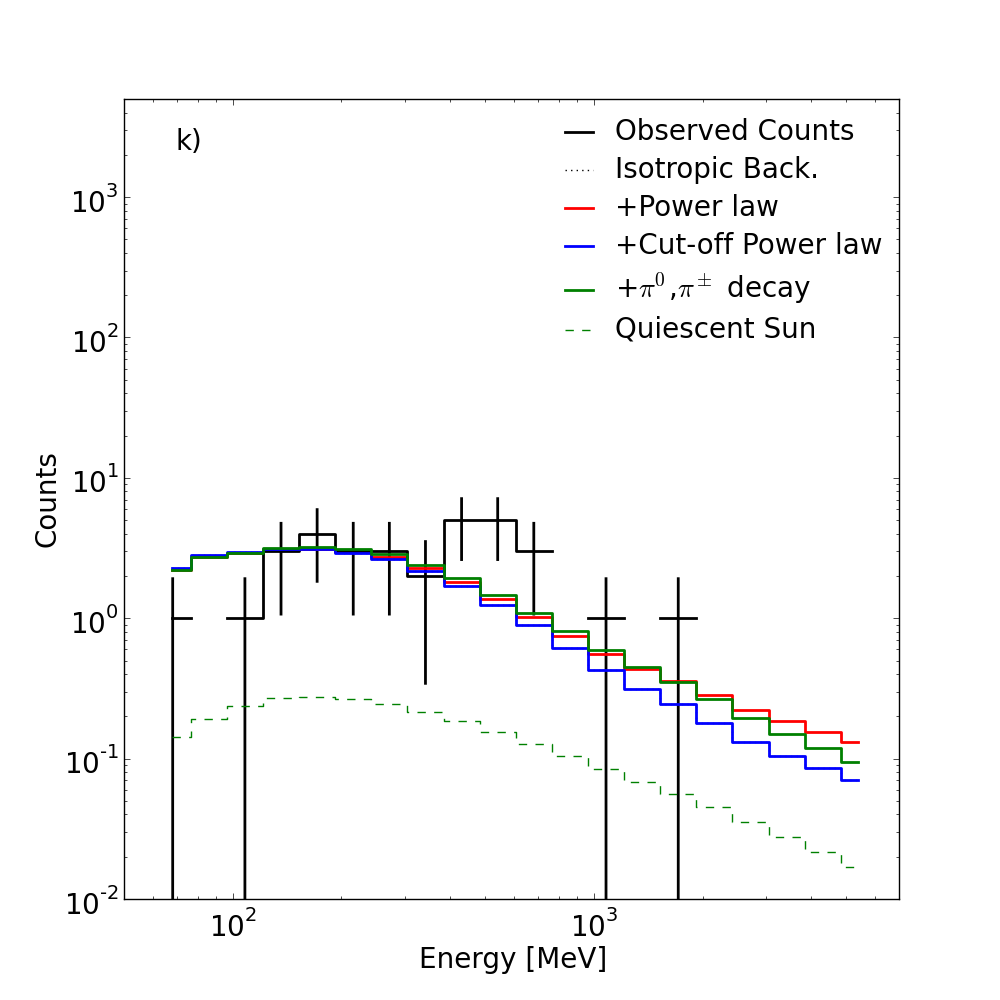}
\caption{Comparison between observed counts (black thick line) and model predictions.
The dotted black line is the isotropic background (sum of the Galactic and isotropic background), the green dashed line is the contribution from the quiescent Sun. The red, green and blue thick lines are the predicted numbers of counts from background + source modeling the solar flare 
with a power law, a  power law with exponential cut-off, and with a pion decay model, respectively. 
Statistical uncertainties are associated to the numbers of observed counts using the \citet{1986ApJ...303..336G} prescription for confidence level in the low counts regime. 
These errors are not considered in the likelihood fit (which only compare the number of observed events with the number of predicted events assuming Poisson statistics) but are useful to visualize the statistical uncertainty due to Poisson fluctuation in each bin. The time intervals are defined in Table \ref{tab:SEDValues}.}
\label{Counts}
\end{center}
\end{figure*}

For each interval, if $\Delta$\,TS $\geq$ 50 then the PLEXP model provides a significantly better fit than the simple power-law and we retain the additional spectral component. 
In these time intervals, we also used the pion decay model to fit the data and estimated the corresponding proton spectral index.
We performed a series of fits with the pion decay template models calculated for a range of proton spectral indices. 
We then fit the resulting profile of the log-likelihood function with a parabola and
determine its minimum ($\like_{\rm min}$) and the corresponding value $s_0$ as the maximum likelihood value of the proton index. 
The 68\% confidence level is evaluated from the intersection of the profile with the horizontal line at $-2$\,$\Delta\log(\like_{\rm min})+1$.
Table \ref{SpectralFit2} summarizes our results.
In Figure~\ref{Counts} we compare the observed count spectra with the predicted numbers of counts for the different models. 
The predicted numbers are the sum of the contribution of the background and of the source, after the spectral parameters are optimized.
The contribution of the isotropic background and of the quiescent Sun is also shown in the figures.
In the first 6 time intervals (\emph{a} through \emph{f}) a power-law model does not correctly reproduce the data, while a curved spectrum (such as the power law with an exponential cut off or a pion decay model) provides a better description of the data.
In the time intervals from g) to j) the power-law representation is sufficient to describe the data; in the last bin, the source is only marginally significant (TS=7).

In the lower panel of Figure~\ref{lightcurve2} we combine the LLE and likelihood analysis results, showing the evolution of  both the  gamma-ray flux and the derived spectral index of the protons\footnote{After approximately 11:00 UTC the flux of the Sun diminished to the point that the spectral index of the proton distribution cannot be significantly constrained.}.
Unlike during the impulsive phase, the spectrum during the temporally extended phase becomes softer ($s$ increases monotonically).
We also compare our results with the GOES proton spectral data. For this, we selected two energy bands ($>$30 MeV and $>$100 MeV) and corrected the light curve by the proton time-of-flight (TOF) to 1 AU by considering the TOF for 30 MeV and 100 MeV protons (i.e. the maximum delay in each energy band).
As a measure of the spectral index of the SEP protons, we compute the Hardness Ratio HR$_{p}$ defined as:
\begin{equation}
{\rm HR}_{p}=\log\frac{\rm P_{>100 MeV}}{\rm P_{>30MeV}}
\end{equation}
from which we calculate the value of the spectral index $s_{\rm SEP}$  of the SEP protons observed at 1 AU that correspond to such $HR_p$ values: 
\begin{equation}
s \sim 1-0.83\,{\rm HR}_{p}
\end{equation}

To estimate the uncertainty associated with this procedure we repeat the calculation neglecting the TOF correction. In this way we obtain two values for the SEP spectral index for each time bin, corresponding to the actual and zero delay due to the time of flight. In Figure~\ref{lightcurve2} we report the estimated proton spectral index as the average of these two values and its uncertainty as half the difference of these two values. Note that the values of the proton spectral index inferred from the gamma-ray data are systematically softer than the value of the index derived directly from SEP observation, although the temporal evolution (hard-to-soft) is similar.

\begin{deluxetable*}{l ccrrccccc}
\tabletypesize{\scriptsize}
\tablecaption{Spectral analysis of the time extended emission}
%Time intervals during which the Sun is in the FoV, and used in our analysis. They are determined by the cuts applied on the data and reflect the particularities of the \fermi-LAT observation capabilities over several hours long events.
\tablewidth{0pt}
\tablehead{ 
\colhead{Interval} & \colhead{Start (UT)} & \colhead{Duration}   	  &\colhead{TS$_{\rm PL}$} & \colhead{$\Delta$TS$^{a}$} & $\Gamma$ & E$_{co}$& \colhead{Flux$^b$} & \colhead{Proton index} & \colhead{X,Y (r$_{68}\oplus70$)$^{c}$}\\
\colhead{}&  \colhead{2012/03/07}   & \colhead{s}   &\colhead{}    & \colhead{}  &\colhead{}    & \colhead{MeV}                    & \colhead{} & \colhead{} &\colhead{arcsec}}
\startdata
a) & 02:27:00 & 1110 & 1400 & 85   	   	& $-$1.1$\pm$0.2 	& 260$\pm$40 & 2.3$\pm$0.8    &  	  3.7$\pm$0.1 & -280, 140 (320)\\ 
b) & 03:52:00 & 2370 & 16421 & 982   	& $-$0.9$\pm$0.1 	& 210$\pm$10 & 3.53$\pm$0.09  & 	4.0$\pm$0.1 &-450, 260 (100)\\ 
c) & 05:38:32 & 1050 & 1393 & 159  	& $-$0.2$\pm$0.3 	& 120$\pm$10 & 3.2$\pm$0.1    & 	4.6$\pm$0.2 & -470, 260 (350)\\ 
d) & 07:03:00 & 2400 & 9003 & 756 		& $-$0.4$\pm$0.1 	& 130$\pm$7 & 2.08$\pm$0.04  &	 4.8$\pm$0.1 & -500, 130 (150)\\ 
e) & 08:50:00 & 1020 & 500 & 73 		& 0.2$\pm$0.6 		& 90$\pm$22 & 1.5$\pm$0.1    & 		5.1$\pm$0.3 & \,670, 580 (750)\\ 
f) & 10:14:32 & 2370 & 1833 & 204 		& 0.3$\pm$0.2 		& 80$\pm$9 & 0.81$\pm$0.03 & 	5.5$\pm$0.2 & \,440, 380 (330)\\ 
g) & 13:25:00 & 2400 & 137 & 13 		& 0.05$\pm$1.0 	& 80$\pm$30 & 0.24$\pm$0.04 &  -- &        --\\ 
h) & 16:36:00 & 780  & 17 & 8	  		& 1.0$\pm$0.1 		& 50$\pm$10 & 0.03$\pm$0.01  & 	-- &        --\\ 
i) & 18:24:00 & 540  & 10 & 3  		& 1.0$\pm$0.1 		& 33$\pm$9 & 0.05$\pm$0.03  & 	-- &        --\\ 
j) & 19:47:00 & 1710 & 59 & 2 			& $-$1.5$\pm$0.8 	& 350$\pm$300 & 0.02$\pm$0.01 & 	-- &        --\\ 
k) & 22:58:30 & 2370 & 7 & 4 			& 1.0$\pm$0.1 		& 230$\pm$70  & 0.010$\pm$0.007& 	-- &        --\\
\enddata
\label{SpectralFit2}
\tablenotetext{a}{$\Delta$TS=TS$_{PLEXP}$-TS$_{PL}$.}
\tablenotetext{b}{Energy Flux of gamma-rays between 100\,MeV and 10\,GeV, in units of $10^{-7}$ erg\,cm$^{-2}$\,s$^{-1}$, calculated using the best fit model.}
\tablenotetext{c}{A systematic error of 70 arcsec has been added in quadrature to the estimated 68\% error radius, and is reported between parenthesis.}
\end{deluxetable*}

%02:27:00 & 1110 & 1328 & 87   	  & 2.3$\pm$0.8    &  	  3.7$\pm$0.1 & -285, 138, 317 \\ 
%03:52:00 & 2370 & 16421 & 982 & 3.53$\pm$0.09  & 	4.0$\pm$0.1 &-445, 259, 101 \\ 
%05:38:32 & 1050 & 1393 & 159  & 3.2$\pm$0.1    & 	4.6$\pm$0.2 & -474, 259, 346 \\ 
%07:03:00 & 2400 & 9003 & 756 	& 2.08$\pm$0.04  &	 4.8$\pm$0.1 & -505, 133, 155\\ 
%08:50:00 & 1020 & 500 & 73 	& 1.5$\pm$0.1    & 		5.1$\pm$0.3 & 672, 581, 745\\ 
%10:14:32 & 2370 & 1833 & 204 	& 0.81$\pm$0.03 & 	5.5$\pm$0.2 & 443, 381, 335\\ 

\begin{figure*}[ht!]
\begin{center}
%\includegraphics[trim=0cm 0cm 2cm 1cm, clip=true, width=1.9in]{plot_NE/plot_00}
%\includegraphics[trim=0cm 0cm 2cm 1cm, clip=true, width=1.9in]{plot_NE/plot_01}
%\includegraphics[trim=0cm 0cm 2cm 1cm, clip=true, width=1.9in]{plot_NE/plot_02}
%\includegraphics[trim=0cm 0cm 2cm 1cm, clip=true, width=1.9in]{plot_NE/plot_03}
%\includegraphics[trim=0cm 0cm 2cm 1cm, clip=true, width=1.9in]{plot_NE/plot_04}
%\includegraphics[trim=0cm 0cm 2cm 1cm, clip=true, width=1.9in]{plot_NE/plot_05}
%\caption{\nob{Photon Spectrum} in the six time bins where the pion decay model provided the best fit to the data. For each time bin we illustrate the models used for fitting the broad band spectrum: power law (dashed), power law with an exponential cutoff (dotted)  and pion template model (solid). 
%In the inserts we report the profile of the likelihood function -2$\Delta\log(\like)$ which is used to estimate the pion template model that best match the data. 
%The scan is done as a function of the index of the proton distribution used to compute the templates. The intersects with the horizontal dashed line represent the 68\% confidence level, used to estimate the errors.}
\includegraphics[trim=0cm 0cm 2cm 1cm, clip=true, width=1.9in]{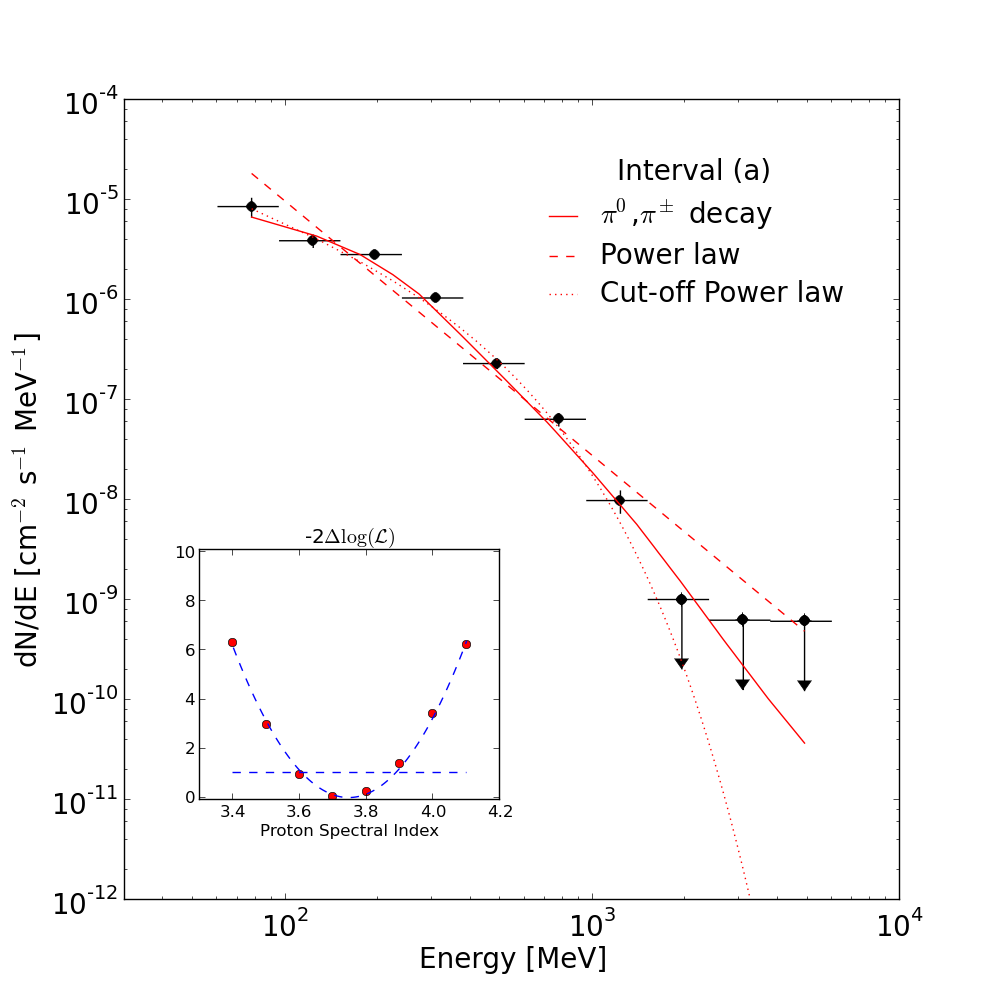}
\includegraphics[trim=0cm 0cm 2cm 1cm, clip=true, width=1.9in]{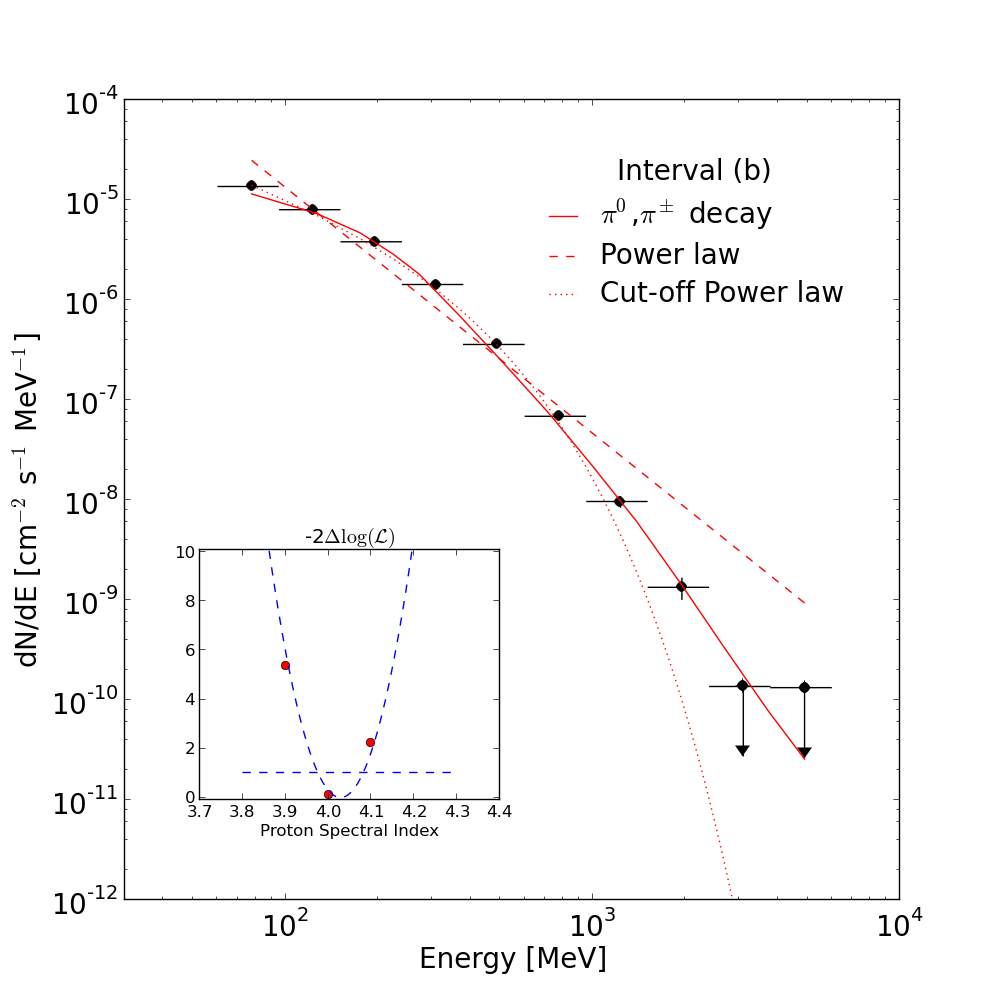}
\includegraphics[trim=0cm 0cm 2cm 1cm, clip=true, width=1.9in]{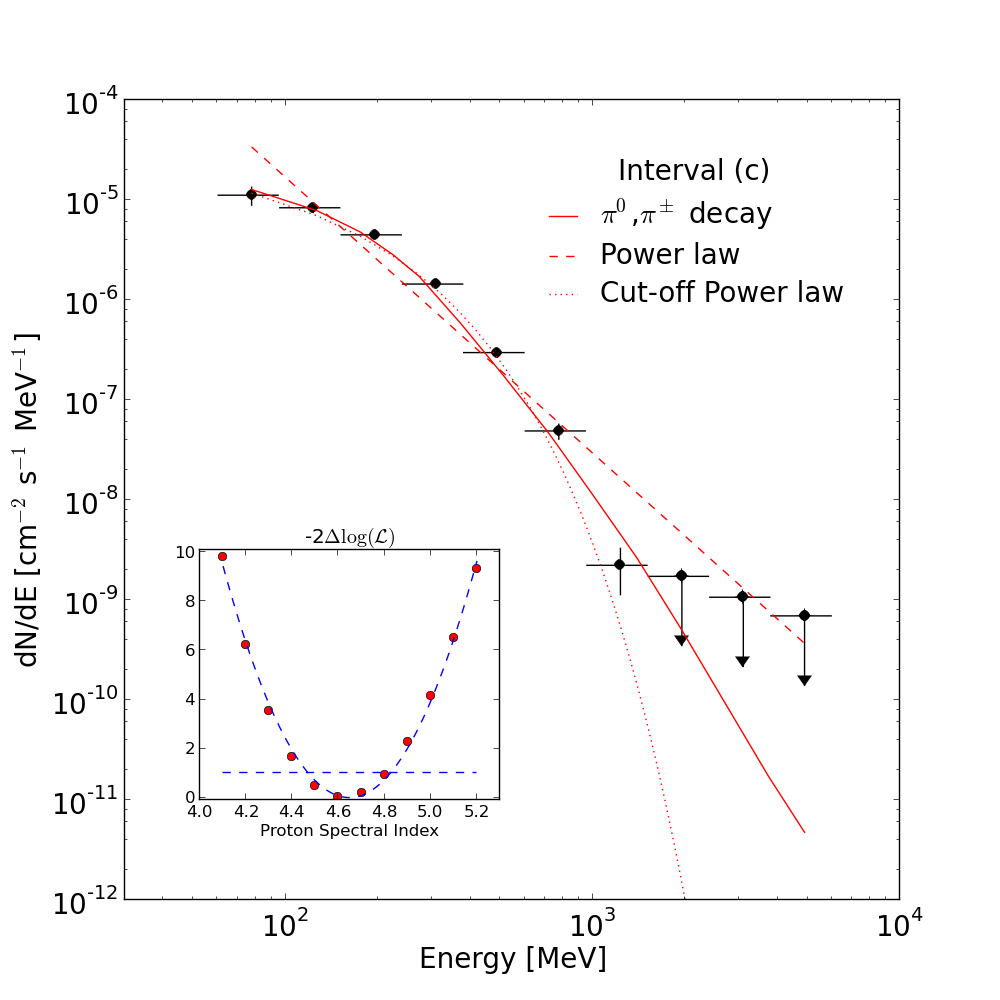}
\includegraphics[trim=0cm 0cm 2cm 1cm, clip=true, width=1.9in]{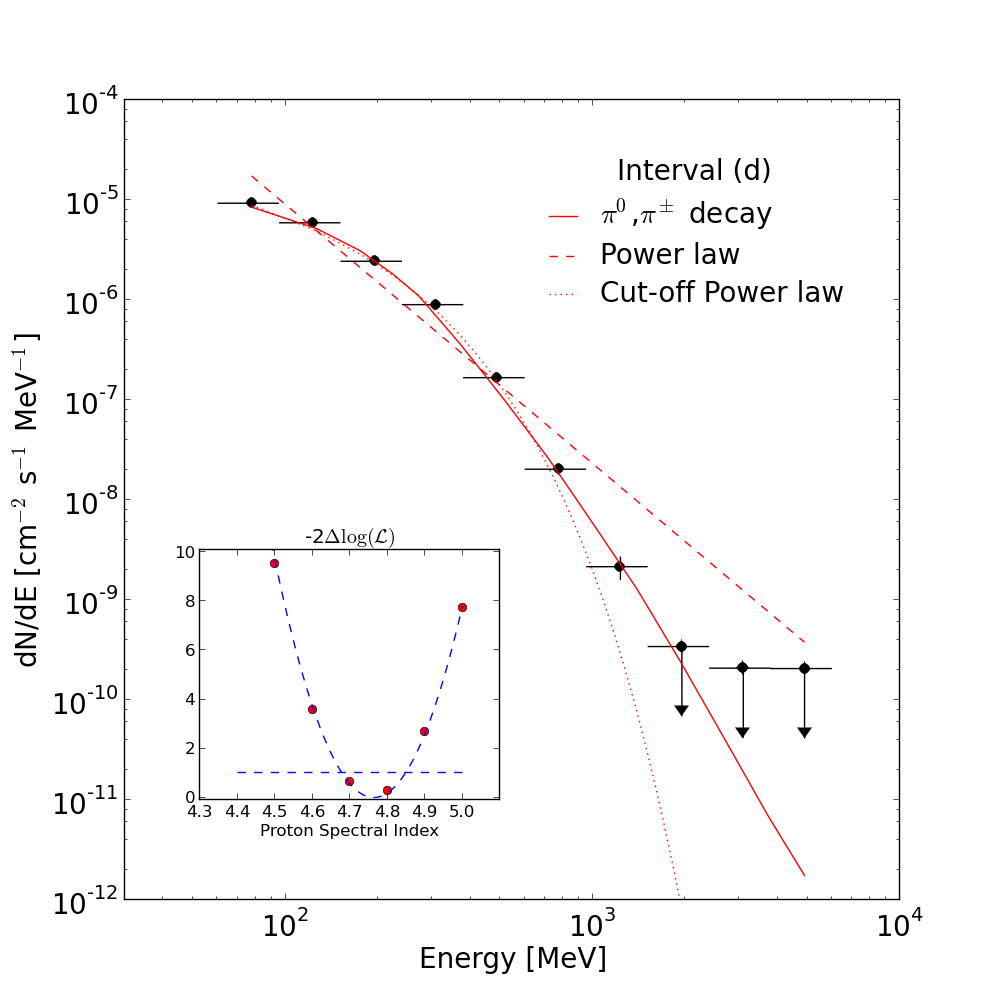}
\includegraphics[trim=0cm 0cm 2cm 1cm, clip=true, width=1.9in]{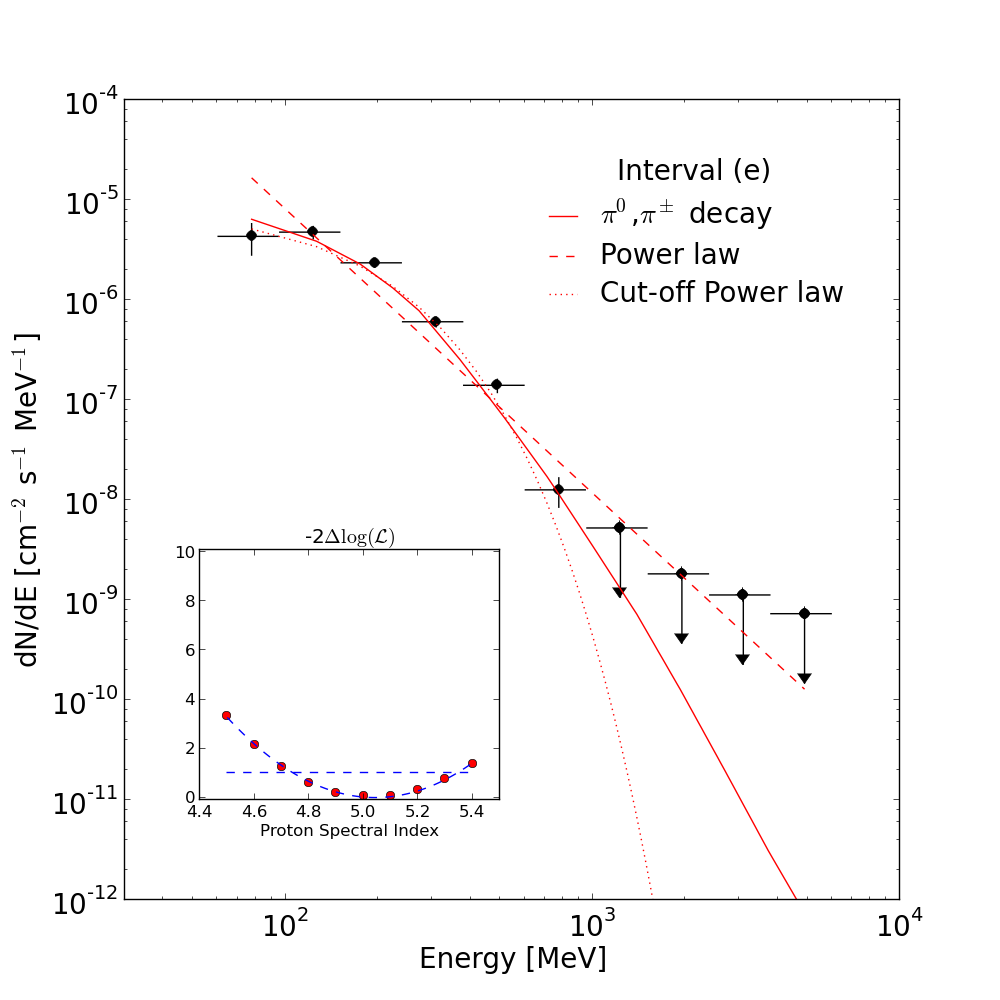}
\includegraphics[trim=0cm 0cm 2cm 1cm, clip=true, width=1.9in]{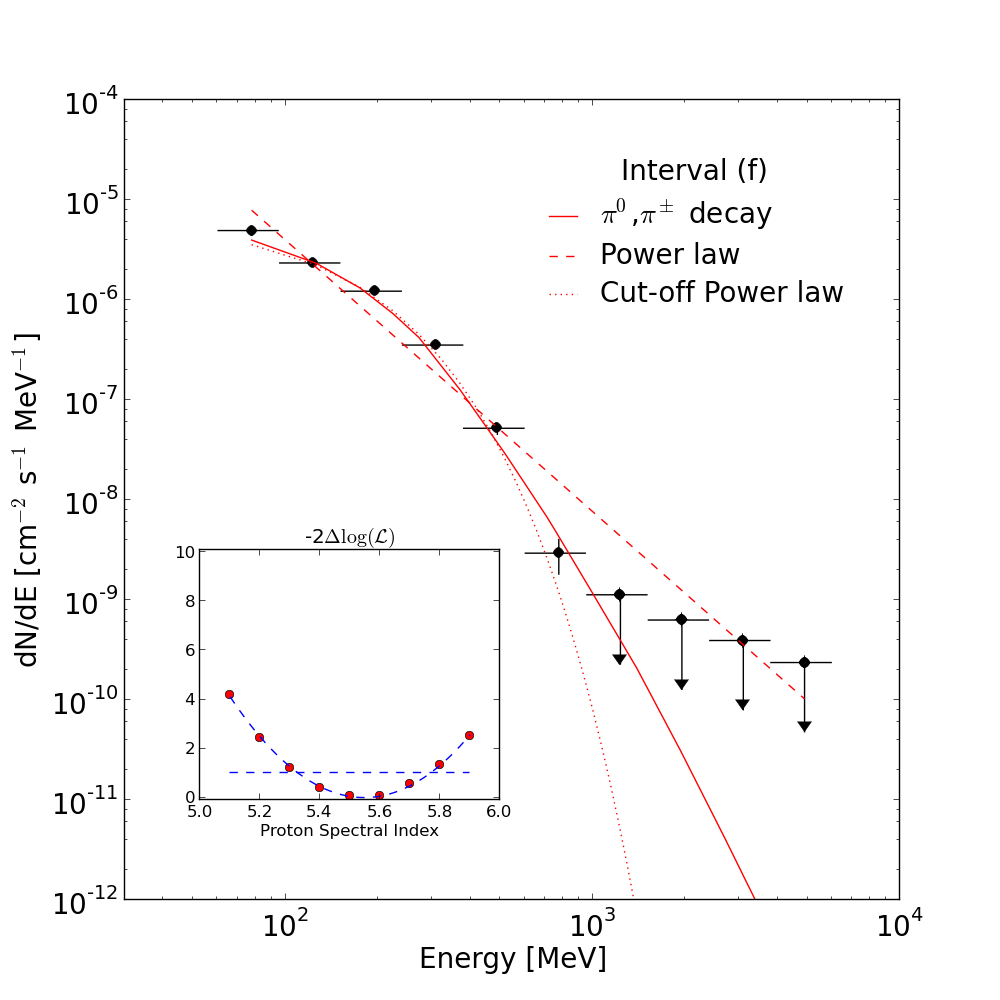}
\caption{Spectral energy distribution in the 6 time intervals for which the pion decay model provided the best fit to the data. For each time interval we illustrate the models used for fitting the broad band spectrum: power law (dashed), power law with an exponential cutoff (dotted)  and pion decay template model (solid). 
In the insets we report the profile of the likelihood function -2$\Delta\log(\like)$ which is used to estimate the pion template model that best match the data. 
The scans are performed as functions of the index of the proton distribution used to compute the templates. The intersections with the horizontal dashed lines represent the 68\% confidence levels used to estimate the errors.}
\label{Spectra}
\end{center}
\end{figure*}

In the 6 time intervals within which the pion decay model provides a better 
fit, we compute the spectral energy distribution by first using the
result of the fit with the power law with an exponential cut-off to 
constrain the background. The normalization of the background is set to the 
best fit value. We divide the data into 10 energy bins and repeat
the spectral analysis in each bin independently. We keep the normalization of
the background constant for the bin-by-bin fits and assume that the in-bin spectrum is an $E^{-2}$ power law, with only the normalization allowed to vary.
For non-detections (TS$<$9), we compute  95\% CL upper limits.
The results are shown in Figure~\ref{Spectra}. We also report the values of the energy flux in the 6 time intervals in Table~\ref{tab:SEDValues}.

%\begin{deluxetable}{cccccccccccc}
%\tabletypesize{\tiny}
%\tablecaption{Spectral Energy Distribution. Energy fluxes in $\times$10$^{-9}$ erg cm$^{-2}$ s$^{-1}$}
%\tablewidth{0pt}
%\tablehead{\colhead{Energy Bin (MeV)} & \colhead{60--95 } & \colhead{95--150} & \colhead{150--239} & \colhead{239--378} & 
%\colhead{378--600} & \colhead{600--950} & \colhead{952--1508} & \colhead{1509--2391} & \colhead{2391--3789} & \colhead{3780--6000}}
%\startdata
%a) & 64.1$\pm$17.0 & 79.3$\pm$13.3 & 148.8$\pm$14.3 & 136.7$\pm$12.7 & 72.9$\pm$9.8 & 48.7$\pm$8.8 & 16.4$\pm$5.9 & $<$13.7 & $<$21.3 & $<$21.9 \\
%\enddata
%\label{SEDValues}
%\tablenotetext{a}{Flux is between 100\,MeV and 10\,GeV, in units of $\times10^{-7}$ erg\,cm$^{-2}$\,s$^{-1}$.}
%\end{deluxetable}

\begin{deluxetable*}{lccccccc}
\tabletypesize{\scriptsize}
\tablewidth{0pt}
\tablecaption{Spectral Energy Distribution}
\tablehead{\colhead{Energy Bin} & \multicolumn{6}{c}{Energy Flux}\\
\colhead{MeV} & & \multicolumn{6}{c}{$\times10^{-9}$\,erg\,cm$^{-2}$ s$^{-1}$}\\
\cline{1-1} \cline{3-8}\\
& & a) & b) & c) & d) & e) & f) }
\startdata
 60--95 &  &  64$\pm$17     		& 117$\pm$6 	& 83$\pm$21 	&	 79$\pm$5 	& 31$\pm$14 	& 40$\pm$5 \\ 
  95--150 & &  79$\pm$13 			& 175$\pm$6 	& 171$\pm$21 	& 129$\pm$5 	& 96$\pm$16 	& 49$\pm$4 \\ 
  150--239 & &  149$\pm$14		& 211$\pm$6 	& 236$\pm$19 	 & 134$\pm$5 	& 121.1$\pm$14 	& 65$\pm$4 \\ 
  239--378 & &  137$\pm$13		& 198$\pm$6 	& 190$\pm$16 	& 123$\pm$5 	& 76.2$\pm$11 	& 46$\pm$4 \\ 
  378--600 & &  73$\pm$10 		& 123$\pm$5 	& 94$\pm$12 	& 56$\pm$4 	& 41.8$\pm$8 	& 16$\pm$3 \\ 
  600--950 & &  49$\pm$8 			& 57$\pm$4.2 	& 36$\pm$8 	& 16$\pm$2 	& 7.5$\pm$4 		& $<$ 4 \\    
  952--1508 & &  16$\pm$6 		& 19$\pm$3 	& $<$ 12 			& 3$\pm$1 		& $<$ 12 			& $<$ 2 \\    
  1509--2391 & &  $<$ 14 			& 6$\pm$2 		& $<$ 10 			& $<$ 2 			& $<$ 10 			& $<$ 4 \\    
 2391--3789 & &  $<$ 21 			& $<$ 5 			& $<$ 15 			 & $<$ 3 			& $<$ 16 			& $<$ 6 \\    
 3780--6000 & &  $<$ 22 			& $<$ 5 			& $<$ 25 			& $<$ 7 			& $<$ 26 			& $<$ 8 \\    
\enddata
\label{tab:SEDValues}
\end{deluxetable*}

\subsection{Localizing the high energy gamma-rays}

\begin{figure*}[ht!]
\begin{center}
\includegraphics[width=2.5in]{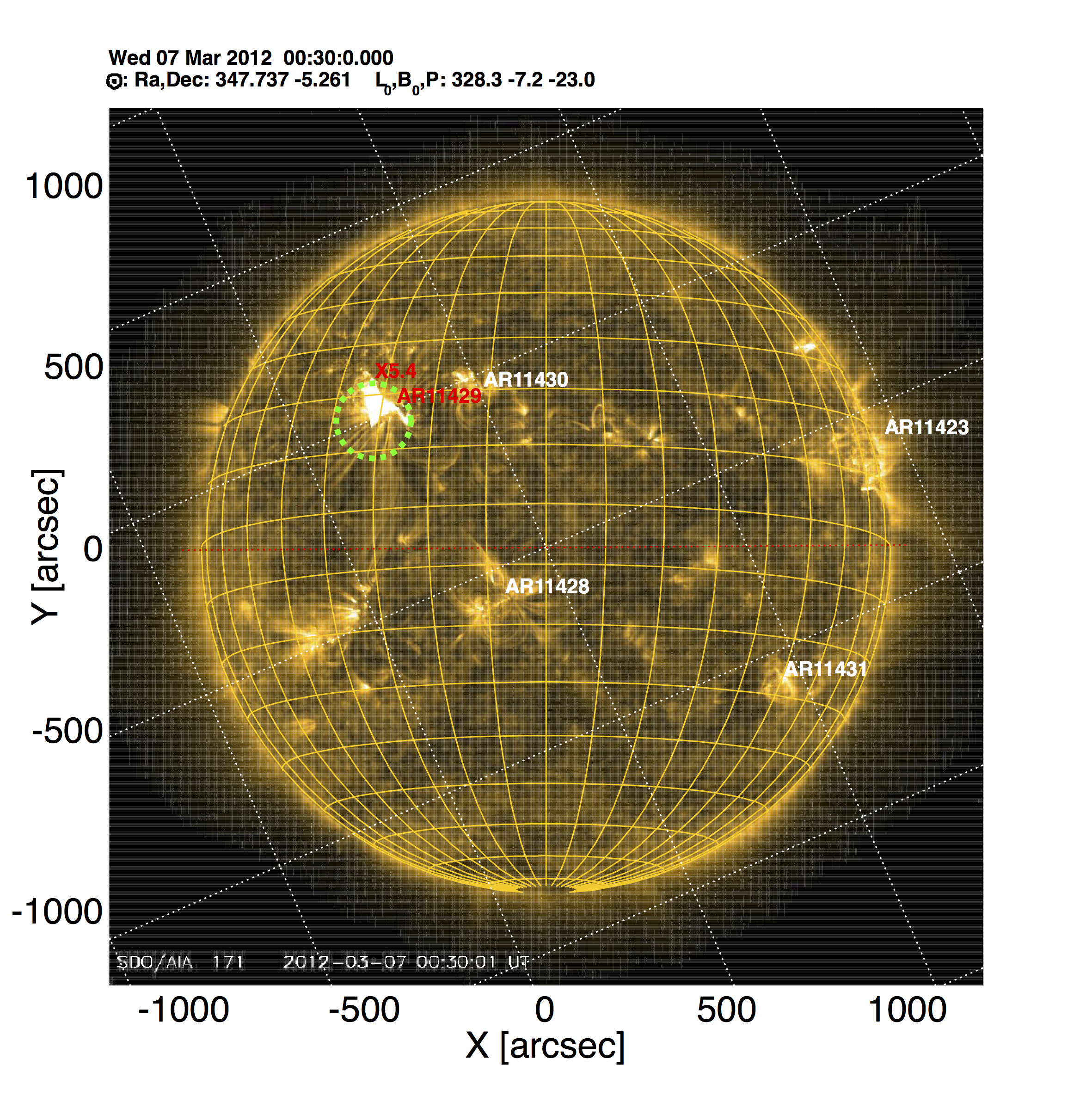}
\includegraphics[width=2.5in]{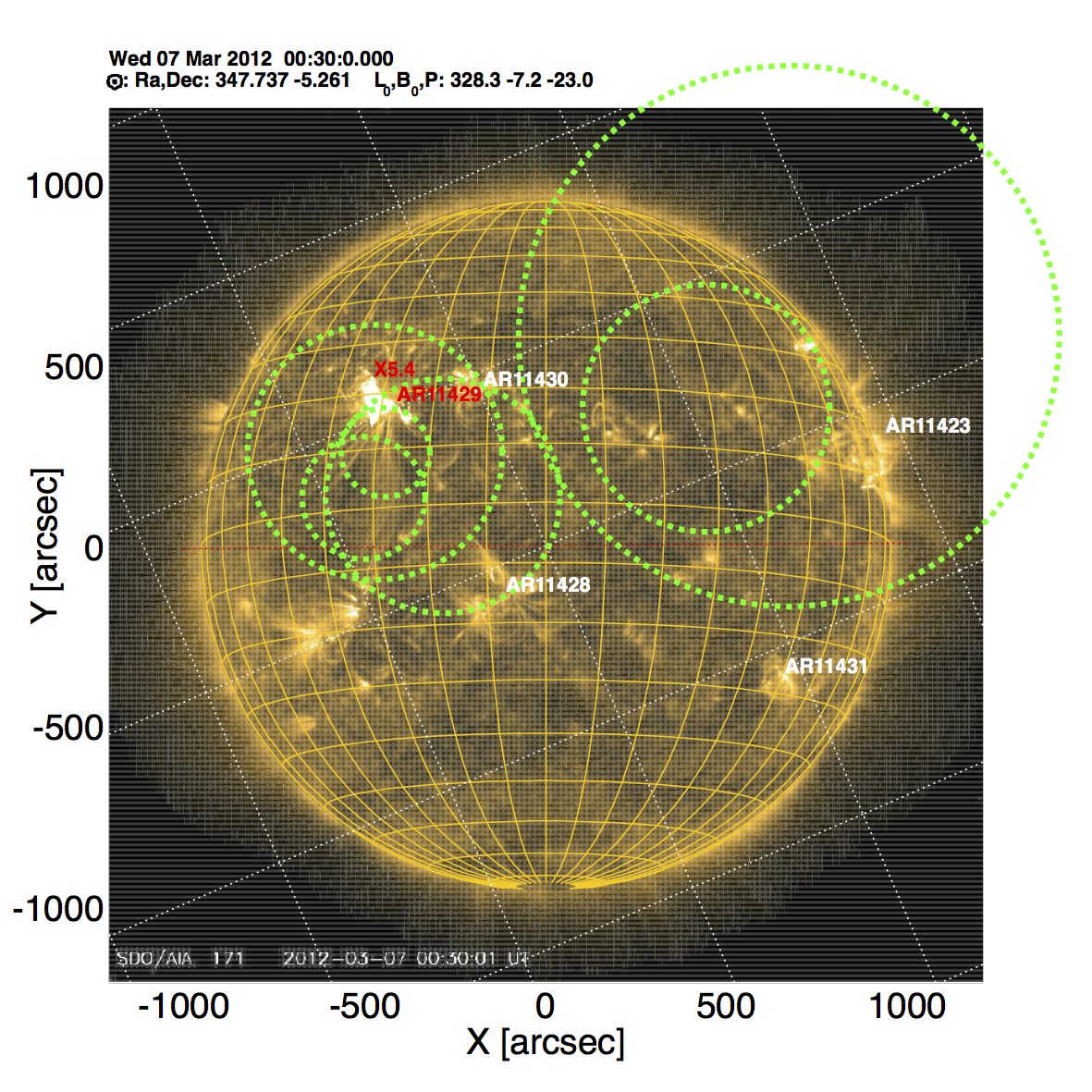}
\caption{Location of the gamma ray emission above 100 MeV for the time-integrated (left) and the time resolved (right) analysis. 
The images on the background are from SDO (AIA 171\AA) and are taken at the time of the flaring episode. Active regions are flagged with their respective NOAA numbers. The region associated with the X-class flares is indicated with a red label, located at N16E30 (X,Y=-471,$373\arcsec$).
The green circles are the 68\% source location uncertainty regions (+systematic error added in quadrature). The grid on the background is the coordinate grid of equatorial coordinates, while the yellow sphere is the heliocentric coordinate grid (with the projected solar rotation axis parallel to the Y-axis, the Z-axis is the line of sight (from the Sun to the observer) and the X-axis in the cartesian projection complete the normal basis.}
\label{SkyMap}
\end{center}
\end{figure*}

We measure the direction of the $>$ 100 MeV gamma-ray emission using the 
\texttt{gtfindsrc} tool and perform a likelihood analysis on 
both time-integrated and in separate time intervals.
The background is modeled using the best-fit parameters obtained by the time-resolved spectral analysis described in the previous section, and the source is modeled according to the best-fit model. The uncertainties on the localization are obtained by combining the 68\% error radius from \texttt{gtfindsrc} with the 
systematic bias associated
with the ``fisheye'' effect in quadrature. We estimate the latter using
Monte Carlo simulations and find it to be 0$\fdg$02 ($\approx$ 70 $\arcsec$).

The results for the 6 time bins where the pion template provides the best fit
are shown in Figure~\ref{SkyMap}. The localization centroids (with uncertainties) are shown in the last column of Table~\ref{SpectralFit1}.
For the remaining time intervals, the reconstructed location of the emission is consistent with the direction of the Sun, 
although the associated uncertainty is larger than the angular diameter of the Sun.

\nob{During the $\sim$ 20 hours of detected flaring gamma-ray emission, the LAT measured 5 photons with E $>$ 2.5~GeV  and reconstructed direction less than 1\de from the center of the solar disk. All 5 of these events belong to the \texttt{P7SOURCE\_V6} event classe and 3 of them are also \texttt{P7ULTRACLEAN\_V6}. Two of these photons, with energies 2.8~GeV and 4~GeV, were detected during the impulsive phase of the flare and the remaining three during the extended emission, including one with  E$=$4.5~GeV at 07:30UT. Comparing the distance from the center of the solar disk and the predicted 68\% containment radius from the point spread function (PSF) of the instrument we find that four of the events are consistent with the solar disk. In the case of the 4.5~GeV photon, the reconstructed direction is 0\fdg8 from the center of the solar disk and the 68\% containment radius is approximately 0\fdg2. Therefore we conclude that the reconstructed direction of this event is only marginally consistent with the solar disk.}

\nob{Considering the average rate of LAT detected photons above $\sim$ 2.5~GeV coming from an ROI with radius 1\de centered at the Sun (calculated using all available flight data, excluding the bright LAT detected solar flare time intervals) we find that the probability to observe five or more events in 8 hours due to Poisson fluctuations is approximately P=8.0$\times10^{-6}$ ($\sim$ 4.8$\sigma$). In Table \ref{HighEnergy} we list some of the basic properties of these photons, including the arrival time, energy, reconstructed distance from the center of the solar disk, reconstructed direction with respect to the instrument coordinate system, 68\% containment radius and conversion type.}

%\nob{The LAT detected few high-energy events. We summarize all the events detected above 2.5 GeV within a reconstructed position of 1\de from the center of the solar disk in Table \ref{HighEnergy} (Need to add a description of the columns).} probability to observe five or more events in 8 hours as Poisson fluctuation from the expected background rate is approximately P=8.0$\times10^{-6}$ ($\sim$ 4.8$\sigma$).Three more photons were detected at later times, including one with  E$=$4.5~GeV at 07:30UT.

\begin{deluxetable*}{ccccccc}
\tabletypesize{\scriptsize}
\tablecaption{High Energy Events}
\tablewidth{0pt}
\tablehead{ 
\colhead{Arrival Time} & \colhead{Energy} & \colhead{Distance}  &\colhead{$\theta$} & \colhead{Event Class} 	&  \colhead{Conversion} 	& \colhead{PSF$_{68\%}^{a}$} \\
\colhead{2012/03/07 UT}                 &  \colhead{GeV}   & \colhead{(deg)}   &\colhead{(deg)}    		& \colhead{}  		&\colhead{}    			& \colhead{(deg)}}
\startdata
0:49 & 2.8 & 0.2 & 49 & SOURCE & FRONT & 0.3 \\ 
1:18 & 4   & 0.6 & 66 & ULTRACLEAN & BACK & 0.5\\ 
2:35 & 2.9 & 0.6 & 62 & SOURCE & BACK & 0.6 \\ 
4:12 & 2.9 & 0.5 & 36 & ULTRACLEAN & BACK & 0.6 \\ 
7:30 & 4.5 & 0.8 & 44 & ULTRACLEAN & FRONT & 0.2 \\ 
\enddata
\tablenotetext{a}{PSF$_{68\%}$ corresponds to the 68\% containment radius calculated from the PSF of the instrument for an energy and direction equal to the energy and direction of the event.}
\label{HighEnergy}
\end{deluxetable*}

%\nob{In particular two events (one ``SOURCE'' and one ``ULTRACLEAN'') of $\sim$2.8 GeV and $\sim$4 GeV were detected at 0:49 and 1:18 UT respectively, still during the first orbit.Considering the average rate of events above $\sim$ 2.5~GeV coming within 1\de from the center of the solar disk (calculated using the full history of the satellite, and excluding period of time during bright LAT detected solar flares, the probability to observe five/four or more events within 8/4 hours as Poisson fluctuation from the expected background rate is approximately P=8.0$\times10^{-6}$/1.1$\times10^{-5}$ ($\sim$ 4.8$\sigma$).

%\input{analysis_multi}
%\input{analysis_temporal}
%\input{analysis_spec}
%\input{analysis_multi_proton}

\def\beq{\begin{equation}}
\def\eeq{\end{equation}}
\def\cc{{\rm cm}^{-3}}

\section{Discussion and Interpretation}
The sensitivity of the \Fermi-LAT enables the investigation of several aspects 
of solar flares that were not previously accessible, in particular the spectral evolution during the impulsive phase
and throughout the temporally extended phase, as well as the localization of the $>100$ MeV emission.
The data for the exceptionally bright solar flares of 2012 March 7 represent an excellent
opportunity to study the details of these characteristics. 
Here we focus on the possibility of 
constraining the emission and acceleration processes.

For the initial four time intervals the projected location of the gamma-ray 
emission is consistent with the position of the active region \#11429. While in the last two time intervals the localizations are slightly displaced with respect to this region, but still consistent with the solar disk. 

GOES fluxes began to rise at about 00:05:00 UT, and continued to increase for 
over an hour, while \Fermi sunrise started roughly six minutes after the peak of the first flare at 00:30:00 UT. This coincided with the gradual decay phase during which the hard X-ray (HXR) emission is relatively soft. The GBM detected only weak emission above 100 keV during the first flare. On the other hand, the second flare had a large flux in the 100-300 keV range and a significant flux above 1 MeV, which indicates acceleration of electrons up to several MeVs. 
The derivative of the GOES flux has a pulse shape with a similar structure to that of the lowest energy GBM channel. 
These pulses show the usual soft-hard-soft spectral evolution in the HXR regime.  
However, the LAT $>100$~MeV emission has a monotonically decreasing flux
that is approximately exponential with a $\sim$~30 m decay period. There is no significant evidence for an upturn in flux during the X1.3 flare, while the derived proton index does show some variation. 
The pion-decay model, which fits well, requires a relatively hard 
proton spectrum with the power-law index, $s$, ranging between $\sim$3.0 and $\sim$3.5. The spectrum is initially soft, but then exhibits evidence of spectral hardening during the second flare, as also appears to be the case in the HXR regime (Figure \ref{lightcurve1}).
The hardening seems to start $\sim$20 minutes before the start of the X1.3 
flare. However, the significance of this early hardening is less than $3\sigma$.
If this is real, explanations for spectral hardening during the decay phase can be an intensification of the acceleration rate or, alternatively, to trapping
of accelerated particles in a coronal loop with a converging magnetic field configuration (see below).

The temporally extended emission is characterized by a slight increase of the
gamma ray flux starting at approximately 2:15:00 UT; the flux reaches its 
maximum at approximately 4:00:00 UT. 
The peak of the light curve is broad and the flux after $t_0$=12:00 UT decays
exponentially as ${ F(t)}\propto\exp\left[-({\rm t-t_{0}})/\tau\right]$, with an exponential decay period $\tau\approx$~2.7 hours.
The gamma ray spectrum and thus the required particle spectrum softens
monotonically during the first six time intervals, with no sign of an early rise or a plateau as seen in the flux. The hardness ratio of the SEP protons also shows similar softening
except there seems to be some deviation at about 10:00 UT which could be due to subsequent events (e.g. the GOES event before 05:00 UT). The SEP proton spectrum is much harder, with index smaller by 2--3 units, than the spectrum of the protons making the gamma rays, as seen in the bottom panel of Figure \ref{lightcurve2}.

We now describe to what extent these new observations constrain the models. In particular, we discuss continuous versus prompt acceleration
in a magnetic trap, proton versus electron emission, and proton versus electron emission and acceleration at the coronal reconnection site versus CME shock.

\subsection{Prompt vs Continuous Acceleration} 
In the prompt acceleration model particles are injected quickly, e.g., as a power-law spectrum, into a trap region where they gradually 
lose energy and emit radiation \citep{murp87,1993A&AS...97..349K}. If the radiation is produced in the trap region then we expect a
spectral variation that depends on the energy dependence of the energy loss rate \citep{asch04}.  For 
relativistic electrons moving in a medium with particle density $n$ and magnetic field $B_{\rm eff}$, the energy-loss rate is given by:
\beq\label{loss}
{\dot E}_L (E)={E_p\over\tau_0}\left[1+\left({E\over E_p}\right)^2\right]
\eeq
where $\tau_0\sim10^4\,(10^{10}\cc/n)$ s  and 
$E_p\sim10\,{\rm MeV}\,(n/10^{10}\cc)^{1/2}(100{\rm G}/{ B_{\rm eff}})$.
This produces characteristic spectra that are flat at low energies ($E<E_p$: due to 
Coulomb energy losses) and have a sharp cut off at high energies ($E>E_p$: due to synchrotron and Inverse Compton (IC) losses) in less than an hour (see e.g. Petrosian 2001)\footnote{Note that even for $B=0$ the IC losses in the optical photon field of the Sun, with effective field of 10 G, makes the energy loss timescale for IC scattering less than a day but not as short as synchrotron losses.}. This clearly disagrees with the data that indicate a power-law spectrum at low energies with a gradual energy cut off.
For $0.1-10$ GeV protons the timescale for the energy loss (due to Coulomb losses),
 $\tau_L$=$(E/{\dot E}_L)\sim 6\times 10^{3}\,(10^{10}\cc/n) (E/0.1\,{\rm GeV})^{1.5}$ s
and is almost constant $\sim 2\times 10^5$ s above 10 GeV ($p$--$p$ interactions). 
This causes a hardening of the spectra (index decreasing by 1.5) within several hours, which also disagrees with the observed spectral evolution. The marginally significant hardening within a few minutes before the X1.5 flare requires a density of $n\gg 10^{10}$ cm$^{-3}$, which is not appropriate for a coronal trap model \citep{1986SoPh..105..365R}.

An alternative scenario is the trap precipitation model \citep{1982AIPC...77..409B} where the trapped particles are 
scattered into the loss cone that causes their precipitation into the chromosphere and below, where they
lose most of their energy and produce gamma rays. Coulomb collisions cannot be the agent for this scattering, because the  
relativistic electron Coulomb scattering rate is lower than the Coulomb energy loss rate by a factor of 
$\gamma^2$ and is much smaller than synchrotron energy loss or IC scattering rate. 
In addition, the Coulomb scattering rate for protons is lower than the energy loss rate for electrons by a factor proportional to the electron to proton mass ratio.
In other words, with Coulomb scattering the particles lose energy before they are scattered into the loss cone. 
Therefore, a much faster scattering rate is required for this scenario. Scattering by turbulence could be a possibility but 
in that case acceleration by turbulence will also be present so we no longer have a prompt model. Thus, we conclude that a more likely scenario is continuous acceleration \citep[e.g. by turbulence; see][]{2004ApJ...610..550P}  with a timescale comparable to, or shorter than, the particle energy-loss timescale.

\subsection{Electron vs proton emission} 
For electrons, non-thermal bremsstrahlung is the only viable mechanism of gamma-ray production \citep{1984AdSpR...4..153T,1987SoPh..111..207V}. However, there are some 
important caveats. The first is that for $>$~100~MeV electrons bremsstrahlung is inefficient. The bremmstrahlung emission time scale $\tau_{\rm brem}\sim 3\times 10 ^4(10^{10}\cc/n)({\rm GeV/E})^{0.1}$~s is much longer than the synchrotron energy-loss time $\tau_{\rm sync}\sim 40\,{\rm(GeV/E)}(100{\rm G}/B)^2$~s and even the IC scattering energy-loss time scale $\tau_{\rm IC}\sim 8\times 10^3 ({\rm GeV}/E)$~s. Thus, a much larger microwave and HXR flux would be expected; whether there are observations that can rule out this possibility is unknown to us. In addition, the highest energy photon observed by the \Fermi-LAT of 4~GeV would require electrons to be accelerated to about 10 GeV. This implies acceleration timescales of less than a few seconds over a period of a day to overcome the aforementioned synchrotron losses. Therefore, protons seem to be more likely agents of the gamma ray production and a power-law spectrum for these particles seems to agree fairly well with the data.
%Whether a simple power law  or a more complicated spectrum is required is beyond the scope of the current paper. 
As mentioned above, protons with energies less than 10 GeV lose energy predominately via Coulomb collisions with the background
electrons with a time scale $\tau_{L}$ that is constant to $> \sim$ 50 hours above several GeV. This indicates that the emission originates from regions with densities much higher than those found in the upper corona. 
This implies thick target emission by protons directed toward the chromosphere that, for a continuous injection spectrum ${\dot Q}(E)$, implies an effective thick target proton spectrum:
\beq
N_{\rm eff}(E) = \left[{\tau_L(E) \over E}\right] \, \int_{E}^{\infty} dE' \,\,{\dot Q}(E').
\eeq
Thus, for an injected power-law ${\dot Q}(E)\propto e^{-\delta}$ the effective spectrum will be a broken power-law steepening with an index change from 1 to 1.5 around several GeV. Whether a  spectrum more complicated than a power law can describe the observations adequately is beyond the scope of the current paper. It should also be noted that the yield of gamma-rays is about 1\% at the pion production threshold of $\sim$ 300 MeV but becomes essentially 50\% above a few GeV (the other half of the proton energy going to neutrinos). 

From the results of the gamma-ray spectral analysis, and using the gamma-ray
yield in \citet{murp87}, we estimate the number and energy of the accelerated protons with kinetic energy $>$30 MeV producing gamma-rays and observed as
SEPs.  %colliding with the Sun is only a small fraction (\nob{$\sim$10\%}) of the particles accelerated as SEP.
During the first impulsive phase the estimated number (energy) of protons
interacting with the Sun is N$_{p}\sim 2.5\times$10$^{33}$ ($\E_{p}\sim$ 2.2$\times$10$^{29}$ erg), while, for the temporally extended emission, is approximately N$_{p}\sim$ 1.0$\times$10$^{34}$ ($\E_{p}\sim$ 6.9$\times$10$^{29}$ erg).
From the GOES observations, we estimate that the number (energy) of SEP protons escaping
the CME shock during the period of time when the gamma-ray flux was high (until March 8) is N$_{\rm SEP}\sim$ 4.0$\times$10$^{34}$ ($\E_{\rm SEP}\sim$ 4.2$\times$10$^{30}$ erg)  or, for the 
full period of time when the proton flux was high (i.e. until approximately 20:00 UT of
March 12) is N$_{\rm SEP}\sim$ 1.37$\times$10$^{35}$ ($\E_{\rm SEP}\sim$ 1.23$\times$10$^{31}$ erg)\footnote{The relative values for protons at other energies will differ from the above numbers because of the differences in indexes.}.
%The total energy of protons $>$100MeV is $\sim$5.0$\times$10$^{28}$ erg and $\sim$5.0$\times$10$^{28}$\nor{Are these numbers supposed to be the same?} for the impulsive and for the long lasting emission respectively, and $\sim$3.1$\times$10$^{30}$ erg from direct integration of the GOES proton flux.
We conclude that protons producing gamma-rays carry significantly less energy than SEP protons observed by GOES.

\subsection{Acceleration at the Corona vs CME Shock} 
Continuous acceleration of protons at the flare reconnection region, whether by stochastic acceleration mechanism~\citep{2004ApJ...610..550P} or by a standing shock can account for most of the spectral observations described above. 
In this model protons escape the acceleration site along closed field lines into the chromosphere and the spectral changes are simply due to the softening of the spectrum as the flare decays. 
In stochastic acceleration by turbulence the accelerated particle spectra become softer as the turbulence weakens, which can naturally explain such a spectral evolution.
Acceleration at the CME shock \citep{2001A&A...378.1046R} is also an attractive explanation because the SEP protons, which are most likely accelerated at the shock and escape in the upstream direction \citep{1990AIPC..203..143R}, show the same kind of spectral evolution. However, these spectra are much harder and have a spectral index similar to the impulsive phase index deduced for protons producing the gamma-rays. In addition, gamma-ray production can occur only in the high-density chromosphere so that CME protons must escape downstream into the highly turbulent region behind the shock and be transported back to the Sun against a high-speed outflow. 
The Large Angle and Spectrometric Coronagraph \citep[LASCO,][]{1995SoPh..162..357B} on board the solar and
Heliospheric Observatory (SOHO) mission, observed a fast CME ejected at
approximately 00:30 UT, 
and measured the speed and acceleration of the head of the fastest segment of
the leading edge \citep{2009EM&P..104..295G}.
The average speed was approximately 2684 km s$^{-1}$, while the acceleration obtained by
fitting the height (vs. time) of the leading edge was $\approx$ $-$88 m s$^{-2}$.
The last available measurement indicates that the speed of the CME at 1 solar radius from the photosphere was
$\approx$2379 km s$^{-1}$.
Using this information, we estimated that in 10 hours, the CME would have
traveled approximately 80 \Rs (0.36 AU), requiring the protons  accelerated by the CME shock front to 
travel back along the magnetic lines connected to the Sun for very large distances.
In this scenario, it is possible that the current sheath (CS) provided a preferred path magnetically connecting the front of the CME shock with the
original flare site.
The displacement of the reconstructed position at later times can be explained 
by larger dispersion of particles due to the longer distance traveled.
This model provides the correct scenario for short acceleration time scales
($\sim$1 hr).
If continuous acceleration happens at the flare site, as
suggested by \citet{2004ApJ...610..550P}, low-energy protons accelerated at the
Sun (vastly larger in number than high-energy protons) can escape along open
field lines, reach the CME, and be re-accelerated as SEPs by the shock front.
This would explain the observed correlation between the gamma-rays and SEP
spectral properties.

Finally, it should be noted that we expect electrons to be accelerated as well. Exactly how 
much energy goes to protons and how much to electrons depends on the acceleration 
mechanism. As shown in \citet{2004ApJ...610..550P} generally more energy goes to electrons 
in more strongly magnetized plasmas like those existing in the solar corona. 
How many electrons will be accelerated and what their radiative signature will be is beyond the scope of this paper. 
However, we note that Nobeyama \citep{1995JApAS..16..437N} radio data at both 17 GHz and 34 GHz show a bright signal starting on March
6 at approximately 22:45 UT, reaching its maximum on March 7 at 01:13, and
ending at approximately 03:02 UT. 
No sign of radio activity is visible at later times, suggesting that
highly energetic electrons might explain part of the $>$100 MeV gamma-ray
emission only at earlier times, whereas the temporally extended emission is likely
to be attributed to energetic protons (and ions).

In summary, in this paper we have presented the analysis of the brightest
solar flare
detected by the \Fermi LAT to date. We have shown that during most of the long
duration emission the gamma-rays appear to come from the same active regions
responsible for the flare emission.
The fluxes and spectra of  the high-energy gamma-rays evolve differently 
during the impulsive phase and the sustained
emission. Also there are correlations and some differences between the fluxes 
and
spectral indexes of the protons required for the production of high-energy
gamma-rays and SEP protons seen at 1 AU.  From these data we suggest that
the most likely scenario for production of high energy gamma-rays is that they are
produced by energetic protons (rather than electrons) that are accelerated in the
corona (rather than in the associated fast CME shock) continuously during the
whole duration of the emission.

%\nor{I think that a conclusion should be kept.}
%In this paper we have presented the analysis of the brightest solar flare
%detected by the \Fermi LAT to date. 
%From the time profile of the high-energy $\g$-ray data and for its spectral
%characteristics, a dichotomy between the impulsive phase and the sustained
%emission is evident.\nor{Our observations also suggest} that the correlation be%tween high-energy
%gamma-rays and SEP seems to be the key for understanding \nor{the particle acceleration mechanisms} and the production of high-energy radiation in solar flares.
%For the first time, gamma rays above a few hundred MeV can be localized with
%good spatial resolution, and their direction, \nor{is} consistent with the solar disk. \nor{Thus suggesting that either the particle acceleration} happens at the flare site,
%even at long time-scales, or \nor{that} particles accelerated at the CME shock need to
%travel back to the Sun, possibly from very large distances. \nor{With the current solar cycle nearing its peak the \Fermi-LAT will continue to play a leading role in the study of the high energy gamma-ray solar flares providing further observations needed to better constrain the particle accelerations models and the orgin of the $\gamma$-ray radiation.} 

\acknowledgements

The \fermi~LAT Collaboration acknowledges generous ongoing support from a number of agencies and institutes that have supported both the development and the operation of the LAT as well as scientific data analysis. These include the National Aeronautics and Space Administration and the Department of Energy in the United States, the Commissariat \`a l'Energie Atomique and the Centre National de la Recherche Scientifique / Institut National de Physique Nucl\'eaire et de Physique des Particules in France, the Agenzia Spaziale Italiana and the Istituto Nazionale di Fisica Nucleare in Italy, the Ministry of Education, Culture, Sports, Science and Technology (MEXT), High Energy Accelerator Research Organization (KEK) and Japan Aerospace Exploration Agency (JAXA) in Japan, and the K.~A.~Wallenberg Foundation, the Swedish Research Council and the Swedish National Space Board in Sweden. \\
Additional support for science analysis during the operations phase is gratefully acknowledged from the Istituto Nazionale di Astrofisica in Italy and the Centre National d'\'Etudes Spatiales in France.

We also wish to acknowledge G. Share for his continuous support and important contribution to the \Fermi LAT collaboration.

\appendix
\section{The LAT Low Energy analysis}
\label{appendix::LLE}
\nob{The LAT Low energy (LLE) technique is an analysis method designed to study bright transient phenomena, such as GRBs and solar flares, in the 30 MeV--1 GeV energy range. The LAT collaboration \citep{LATPaper} developed this analysis using a different approach than the one used in the standard photon analysis which is based on sophisticated classification procedures \citep[a detailed description of the standard analysis can be found in][]{LATPaper,2012ApJS..203....4A}. The idea behind LLE is to maximize the effective area below $\sim$ 1 GeV by relaxing the standard analysis requirement on background rejection. The basic LLE selection is based on a few simple requirements on the event topology in the three subdetectors of the LAT namely: a tracker/converter (TKR) composed of 18 x--y silicon strip detector planes interleaved with tungsten foils; an 8.6 radiation length imaging calorimeter (CAL) made with CsI(Tl) scintillation crystals; and an Anticoincidence Detector (ACD) composed of 89 plastic scintillator tiles that surrounds the TKR and serves to reject the cosmic-ray background.}

\nob{First of all, an event passing the LLE selection must have at least one reconstructed track in the TKR and therefore an estimate of the direction of the incoming photon. Secondly, we require that the reconstructed energy of the event be nonzero. The trigger and data acquisition system of the LAT is programmed to select the most likely gamma-ray candidate events to telemeter to the ground. The onboard trigger collects information from all three subsystems and, if certain conditions are satisfied, the entire LAT is read out and the event is sent to the ground. We use the information provided by the onboard trigger in LLE to efficiently select events which are gamma-ray-like. In order to reduce the amount of photons originating from the Earth limb in our LLE sample we also include a cut on the reconstructed event zenith angle (i.e. angle $<$90$^{\circ}$). Finally we explicitly include in the selection a cut on the region of interest, i.e. the position in the sky of the transient source we are observing. In other words, the localization of the source is embedded in the event selection and therefore for a given analysis the LLE data are tailored to a particular location in the sky.}

\subsection{LLE response files}
The LLE response files are generated based on dedicated Monte Carlo simulations. The simulations are used to study how the detector is ``illuminated'' by a source of a known flux and known position, during the real pointing history of the LAT. 
We do this by simulating a bright point source with a spectrum $dN/dE\sim E^{-1}$ at the position of the source in question (the Sun in this case), and using the pointing information saved in the spacecraft data file (\texttt{FT2} file). We use the Fermi-LAT full simulator \citep{2006NuPhS.150...62B} to generate particles from the point source, incident over a cross-sectional area of 6 m$^2$, which illuminates the entire LAT. 
The LAT detector is represented by a complex model containing more than 34000 volumes. Gamma-ray conversion and particle propagation through the detector is implemented using \texttt{GEANT4} \citep{2003NIMPA.506..250G} while digitization and reconstruction are done using the same algorithm used for flight data.
We then apply the LLE selection and bin the resulting events in reconstructed energy versus Monte Carlo energy (\texttt{McEnergy}) obtaining the so-called Redistribution Matrix, $R_{ij}$. This matrix is proportional to the probability that an incoming photon of energy $E\in[E_{j},E_{j+1}$] will be detected in the reconstructed energy bin [$E_{i},E_{i+1}$]. We re-normalize each bin such that: 
\begin{equation}
\sum_{i}R_{ij}=A_{j}=6\times 10^{4} {\rm cm}^2 \fraz{N_{j}({E}_{j+1}-E_{j})}{N_{\rm TOT}({E}_{MAX}-{E}_{MIN})},
\end{equation}
where $N_{\rm TOT}$ is the total number of simulated events over an area of 6\,m$^2$ (typically 10$^7$), 
$N_{j}$ is the number of detected events (that survive the selection cuts) with a \texttt{McEnergy} between $E_{j}$ and $E_{j+1}$.  $A_{j}$ is usually defined as the effective collecting area of the instrument.
The Redistribution Matrix File (RMF) is saved in the standard \texttt{HEASARC RMF} File Format\footnote{Described here: \url{http://heasarc.gsfc.nasa.gov/docs/heasarc/caldb/docs/memos/cal_gen_92_002/cal_gen_92_002.html\#Sec:RMF-format}.}
%$R_{ij}$ is often referred to as the Redistribution Matrix because it describes how a photon of energy E$_{j+1}<$ E$<$E$_{j}$ is ``redistributed" into output detector channels.
\subsection{Orbital background subtraction}
In the case of short and bright transient gamma-ray sources, it is possible to select time windows before and after the transients (the ``off-pulse'' region) and, excluding the time window of the transient itself, fit the count rates in the off-pulse region with a polynomial function and in this way estimate, the background in the time interval of the transient.
For LLE data, this analysis is described in \citet{PelassaLLE} and was applied to the June 12 2010 flare, as presented in \citet{2012ApJ...745..144A}.
This approach relies on a few assumptions: the background should not vary too much during the transient emission and also the amount of statistics should be 
sufficient to constrain the fit. 
These assumptions usually are satisfied when the interval of the transient emission is shorter than the \Fermi orbital period ($\approx$ 90 minutes).

For the 2012 March 7 flare presented here, the standard LLE approach to estimate the
background by fitting the intervals before and after the flare could not be 
used because the flaring episode lasted longer than the
orbital period. \nob{Instead, we estimate the background using data acquired in other orbits.} \Fermi passes through approximately the same geomagnetic configuration every 15 orbits, but given the standard rocking profile (alternating one orbit north and one orbit south) only the 30$^{th}$ orbit approximates similar geomagnetic and pointing conditions, and consequently a similar background rate. 
We do not average multiples of 30 orbits because they become less reliable, as they span observations further removed in time. This method of background estimation has been used for a number of background-dominated instruments in solar flare analyses, including historically for SMM-GRS \citep{1987ApJ...322.1010V,1990ApJ...358..298M} and EGRET \citet{kanb93}\footnote{In \citet{kanb93} an average for $\pm$N$\times$16 orbits (for N=1,2,3) was used.} and currently for Fermi GBM  \citep{Fitzpatrick:2011sf}.%Note that this method is conceptually equivalent to the one applied to EGRET data, although \citet{kanb93} used an average for $\pm$N$\times$16 orbits (for N=1,2,3).The GBM Collaboration has also implemented a similar method for the analysis of GBM data \citep{Fitzpatrick:2011sf}.  

The background files produced for each interval are saved as standard \texttt{PHA-I}\footnote{\url{http://heasarc.gsfc.nasa.gov/docs/heasarc/ofwg/docs/spectra/ogip_92_007/node5.html}}  files. We used \texttt{XSPEC} to execute a forward-folding fit, where the model $M(p_1,p_2,...)$ is folded with the redistribution matrix,$R_{ij}$, and the results are compared with the background-subtracted signal $S_{i}$:
\begin{equation}
S_{i}=C_{i}-B_{i} \stackrel{?}{=} \sum_{j}R_{ij}{M}_{j}({p}_1,{p}_2,...);
\end{equation}
where $M_{j}(p_1,p_2,...)$ is the expected number of events between $E_{j}$ and $E_{j+1}$ for the time interval being analyzed.
A maximum likelihood algorithm is then used to calculate the set of parameters that best model the data (see the \texttt{XSPEC} manual for details.)

\subsection{Validation and systematic uncertainties}

A detailed paper on the assessment of the systematic errors is in preparation; here we summarize the main results. 
Generally speaking, discrepancies between the actual response of the LAT and the response matrix derived from simulations can cause systematic errors in spectral fitting. We investigated the systematic uncertainties tied to the LLE selection by following the procedure described in \citet{LATperform}. In particular, we compared Monte Carlo with flight data, using the Vela pulsar (PSR J0835--4510) as a calibration source. The pulsed nature of the gamma-ray emission from this source \citep{thompson75} gives us an independent control on the residual charged particle background. In fact, off-pulse gamma-ray emission is almost entirely absent, and a sample of ``pure photons'' can be simply extracted from the on-pulse region, after the off-pulse background is subtracted. %The data sample obtained from the \texttt{DIAGNOSTIC} filter and from the event selection described above allowed us to probe the gamma-ray efficiency of the \texttt{GAMMA} filter in different energy bins and for different off-axis angle. 
Considering all time intervals during which the Vela pulsar was observed at an incidence angle $\theta$$<$80$^\circ$, we estimate the discrepancy between the efficiency of the selection criteria in the LAT data and in Monte Carlo to be $\sim$17\% below 100~MeV, decreasing down $\sim$8\% at higher energies, with an average value $\sim$9\% \nob{(note that this average is weighted by the Vela spectrum)}. 

Additionally, we performed a spectral analysis of the Vela pulsar, comparing LLE results with standard likelihood analysis. The \nob{$>$100 MeV} flux obtained from the LLE analysis is 20\% lower than published in \citet{LATVela} and 16\% less than the flux reported by \citet{LATVela2}. In our analysis of the impulsive phase of the 2012 March 7 flare we added a conservative 20\% systematic error in quadrature (represented by the grey band in Figure \ref{lightcurve1}). 

Finally, we also studied the energy resolution using large samples of simulated events with the Fermi-LAT full simulator. No significant bias was found, and the energy resolution for LLE is estimated to be $\sim$ 40\% at 30 MeV, $\sim$30\% at 100 MeV and $<$ 15\% above.

%starting from the efficiency of the \texttt{GAMMA} filter, of the tracker finding algorithm, up to the final event selection that ultimately defines the effective area. We investigate two possible sources of systematic errors. 

%In LLE analysis, we use \texttt{EvtEnergyCorr} as energy estimator, which has the advantage of being defined for most of the events.
%It is based on energy depositions in both the tracker and the calorimeter although, for events with energy below 100 MeV, the energy deposited in the tracker is the main contribution.

\subsection{File Format and Availability}

LLE data are generated for each burst (GRB or solar flare) that trigger the GBM. 
\nob{We first bin the data in energy and time, and, following the procedure described in \citep{LATGRBCatalog}, we select the background region by selecting all the LLE events before the trigger and after 300 seconds from the trigger. We fit each energy bin with a polynomial function of the cosine of the source bore-sight angle as a function of time and we interpolate the background fit into the signal region. We evaluate statistical fluctuation of the signal above the expected background. This procedure is optimized taking into account different signal region and different time binning. We finally look at the post trial probability.} Every detected signal \nob{with a post trial probability greater than 4$\sigma$} is promptly made available through the {\tt HEASARC} web site\footnote{FERMILLE, at \url{http://heasarc.gsfc.nasa.gov/W3Browse/fermi/fermille.html}}.
For each such detected GRB or solar flare, six different files are delivered:
\begin{enumerate}
\item The {\tt LLE event file} format is similar to the LAT photon file format with some exceptions. Because the LLE data are tightly connected to a particular object (position and time), the {\tt FITS} keyword {\tt OBJECT} has been added to the file. Generally, {\tt OBJECT} will correspond to the entry of the GBM Trigger Catalog\footnote{\url{http://heasarc.gsfc.nasa.gov/W3Browse/fermi/fermigtrig.html}} used to generate LLE data and corresponds to the ``name'' column in the {\tt FERMILLE} table (and in the GBM Trigger Catalog table). The direction of the source used for selecting the data for the LLE file is also written in the header of each extension of each LLE file. {\tt PROC\_VER} corresponds to the iteration of the analysis of LLE data. {\tt PASS\_VER} corresponds to the iteration for the reconstruction and the general event classification (Pass6, Pass7, etc.). {\tt VERSION} corresponds to the version of the LLE product for the particular GRB or solar flare represented in the file.

\item The {\tt CSPEC} file is obtained from directly binning the LLE event files. It provides a series of spectra, accumulated with 1 s binning (typically from $-$1000 to 1000 s around the burst). 
Each spectrum is binned in 50 energy channels, ranging typically from 10 MeV to 100 GeV. The format of the {\tt CSPEC} file is tailored to satisfy {\tt rmfit}\footnote{\url{http://fermi.gsfc.nasa.gov/ssc/data/analysis/user/}} standards, and it is not directly usable in {\tt XSPEC}.

\item The {\tt CSPEC Response} file (the RSP file) is the detector response matrix calculated from Monte Carlo simulation, and it corresponds to a single response matrix for each GRB or solar flare.

\item The {\tt PHA-I}file contains the count spectrum. The {\tt PHA-I} file is created from the same time interval used to compute the response matrix.

\item The selected events file is identical to the LLE event file with an additional selection on time interval applied to match the selection used to compute the response matrix and {\tt PHA-I} files.

\item The LAT pointing and livetime history file is identical to the standard LAT file but with entries every s (instead of every 30 s). It typically spans the range $\pm$4600 s from the trigger time.
\end{enumerate}

The complete LLE selection used to select the events is saved in the keyword {\tt LLECUT} in the primary header of each LLE file. If the GBM catalog position of the burst is updated (due to a refined localization from the LAT or Swift or from subsequent on-ground analysis), the LLE data are automatically updated and new versions of the LLE files are produced. In some cases, LLE data are manually generated (using a better localization which may or may not have been used in the GBM Trigger Catalog). If the direction of a GRB is revised based on follow-up observations with other instruments, regenerated LLE files will have the {\tt VERSION} number incremented, but will leave the {\tt PASS\_VER} and {\tt PROC\_VER} unchanged.

In general we do not deliver the background estimates for the time ranges around the burst triggers, and we let the user estimate the background using the procedure described in \citet{PelassaLLE}. 
For the reason explained above, we cannot perform this analysis on the 2012 March 7 flare. Therefore, for this particular flare, we provided LLE data including background files.

%%%%%%%%%%%%%%%%%%%%%%%%%%%%%%%%%%%%%%%%%%%%%%%%%%
% Bibliography
%%%%%%%%%%%%%%%%%%%%%%%%%%%%%%%%%%%%%%%%%%%%%%%%%%
%
\bibliography{SOL120307}
\bibliographystyle{apj}
\clearpage
\end{document}